\documentclass[runningheads]{llncs}
\usepackage[T1]{fontenc}
\usepackage{graphicx}
\usepackage{amsmath,amsfonts}

\usepackage{algorithmic}
\usepackage{array}
\usepackage{textcomp}
\usepackage{stfloats}
\usepackage{url}
\usepackage{verbatim}
\usepackage{booktabs}
\usepackage{multirow}
\usepackage{array}
\usepackage{todonotes}
\usepackage{tikz}
\usepackage{subcaption}
\usepackage{tabularx}
\usepackage{placeins}
\usepackage{hyperref}
\usepackage{url} 
\usepackage{float}
\usepackage{longtable}
\usepackage{xspace}
\usepackage{makecell}

\newcommand{\paragraphB}[1]{\vskip 4pt \noindent \textbf{#1.}}

\newcommand{\sys}{InstantCoFFEA\xspace}
\begin{document}
\title{InstantCoFFEA: Fast and accurate Tor flow correlation at realistic scale}
%
%
\author{Saeyeon Hong\inst{1}, Hyewon Kim\inst{1}, Mohammad Saidur Rahman\inst{2}, Se Eun Oh\inst{1}}

%
\authorrunning{Anonymous Author(s)}
\authorrunning{S. Hong et al.}

%
\institute{Ewha Womans University, \\
\email{\{saeyeonhong, hhongyeahh, seoh\}@ewha.ac.kr}\\
\and
The University of Texas at El Paso\\
\email{\{msrahman3\}@utep.edu}\\
}
%
\maketitle              
\begin{abstract}
Tor end-to-end correlation attacks can link a client to its destination by matching traffic observed at both ends of a Tor circuit. However, their practical threat remains unclear for two reasons: existing correlation methods rely on $O(N^2)$ pairwise processing, and prior works do not reflect the flow counts and base rates seen by realistic adversaries. 
We present \sys, which replaces exhaustive pairwise search with approximate nearest neighbor (ANN) retrieval using CAGRA and introduces ranked voting. At $N\approx200$k, \sys is $14.56\times$ faster than DeepCoFFEA and improves PR-AUC from 0.94 to 0.96. 
We also use TorPS to estimate the flow counts and matching-pair base rates observed by relay-level and network-level adversaries. Our simulations show that realistic attacks involve substantially more candidate flows and much lower base rates than those used in prior evaluations. Under a low-base-rate setting, PR-AUC drops for all methods, but InstantCoFFEA achieves 0.53 compared with 0.42 for DeepCoFFEA. These results show that evaluating correlation attacks requires considering both computational scale and the sparsity of true matching pairs.

\end{abstract}

\section{Introduction}
\label{sec:intro}
Tor~\cite{tor} is a low-latency anonymity network that, as of May 2026, consists of approximately 10,000 relays and serves about 3 million users per day~\cite{tormetrics}. A Tor client builds a \emph{circuit}, typically consisting of an entry guard, a middle relay, and an exit relay. Traffic is forwarded through this circuit using layered encryption, so no single relay learns both the client and its destination.

We use \emph{flow} to refer to the packet sequence observed at one side of Tor for a network connection or session. A flow observed at the entry-guard side is an \emph{egress flow}, while one observed at the exit-relay side is an \emph{ingress flow}. Although Tor encrypts packet contents, metadata such as packet timing and size remains observable. An adversary that observes both sides can use this metadata to match an egress flow with its corresponding ingress flow and link the client to the destination~\cite{Tor_Onion_router}.

Earlier studies considered such attacks difficult to carry out at scale. Later work, however, showed that an adversary does not need to observe the entire Tor network to compromise user paths~\cite{feamster&Ding,users_get_routed,DeNASA,raptor}. This motivated a line of work~\cite{deepcoffea,deepcorr} on accurately correlating noisy traffic, where timing jitter, congestion, and network delay make the two ends of a connection similar but not identical.

DeepCorr~\cite{deepcorr} uses a neural network to score pairs of flows, but correlating $N$ flows on each side requires $N^2$ neural-network evaluations. DeepCoFFEA~\cite{deepcoffea} reduces this cost by mapping flows into an embedding space, requiring only $2N$ neural-network evaluations. Its correlation stage, however, still processes all $N^2$ egress--ingress pairs. This includes similarity computation, sorting, and voting.

One missing piece is the actual scale of the attack. Prior work evaluates correlation at fixed values of $N$, but it is unclear how many flows a relay-level or network-level adversary would observe in practice. This matters because the cost of DeepCoFFEA grows with the number of candidate pairs.

Another issue is the number of true matches. Most observed egress and ingress flows do not correspond to each other. We define the \emph{base rate} as the fraction of all candidate egress--ingress pairs that are true matching pairs. Prior studies~\cite{raptor,deepcorr,deepcoffea} evaluate at much higher base rates than those we observe in our simulation. When the base rate is low, even a small number of false positives can have a large effect on precision.

We first address the computational cost with \sys. As in DeepCoFFEA, each flow is divided into overlapping time \emph{windows}, and each window is mapped to an embedding. \sys uses CAGRA~\cite{CAGRA} to index the ingress embeddings and retrieves only the top-$k$ candidates for each egress-flow window. This avoids exhaustive $N^2$ search, removes the need to sort all ingress candidates, and limits voting to the returned neighbors. At $N\approx200$k, \sys is $14.56\times$ faster than DeepCoFFEA. It also improves the area under precision-recall curve (PR-AUC) from $0.94$ to $0.96$.

We then use TorPS~\cite{users_get_routed} to estimate the flow counts and base rates seen by realistic adversaries. The simulation uses current Tor consensus information, circuit volumes from prior measurement studies~\cite{safely_measure,understanding_tor_usage}, destination distributions from Tranco and Citizen Lab~\cite{tranco_list,citizen_lab}, and network paths for relay-level and network-level adversaries.

The simulated attack scale is much larger and sparser than the settings used in prior correlation studies~\cite{raptor,deepcorr,deepcoffea}. In one relay-adversary setting, the adversary observes 13,889 guard-side flows and 36,300 exit-side flows, but only 352 pairs are true matches. The corresponding base rate is $6.98\times10^{-5}\%$. At larger network-adversary scales, DeepCoFFEA can run out of GPU memory when constructing the full similarity matrix, while \sys remains runnable. Under the low-base-rate relay setting, \sys achieves a PR-AUC of $0.53$, compared with $0.42$ for DeepCoFFEA.

Our contributions are as follows:
\begin{itemize}
\item We identify the $N^2$ correlation stage of DeepCoFFEA -- similarity computation, sorting, and voting -- as its main computational bottleneck, and replace exhaustive pairwise search with ANN retrieval.
\item We estimate realistic flow counts and base rates for relay-level and network-level adversaries using Tor simulation.
\item We evaluate flow correlation at these scales and base rates and show how both computation cost and sparse true matches affect attack performance.
\end{itemize}

\section{Background and Related Work}
\label{sec:background}


\paragraphB{End-to-end Tor Flow Correlation}
Tor clients construct circuits using three relays---a guard, middle, and exit---selected based on network information published in the consensus document. Although Tor encrypts packet contents, its original design acknowledges vulnerability to traffic correlation~\cite{Tor_Onion_router}. An adversary observing both the guard and exit sides can use metadata such as packet timing and traffic volume to link a client to its destination. Prior work has studied this threat under several adversary models, including AS-level observers~\cite{feamster&Ding}, IXPs~\cite{IXP_adversary}, and partial-network observers~\cite{murdoch_and_danezis}. Later studies considered broader network visibility~\cite{as_awareness} and routing effects such as asymmetric BGP paths~\cite{raptor}. Johnson \textit{et al.} developed TorPS~\cite{users_get_routed} to simulate Tor path selection and compromise; we extend TorPS in Section~\ref{sec:evaluation} to estimate the flow counts seen by different adversaries.

To identify matching flows, early correlation techniques used statistical measures such as mutual information~\cite{mutual_info1,mutual_info2}, Pearson correlation~\cite{pearson1,pearson2}, and cosine similarity~\cite{cta}. DeepCorr~\cite{deepcorr} later used a convolutional neural network (CNN) to learn the correlation metric directly from data. Since each pair of flows is processed by the network, correlating $N$ flows on each side requires $O(N^2)$ neural-network evaluations. DeepCoFFEA~\cite{deepcoffea} reduces this cost using feature embedding networks (FENs), which map flows into an embedding space before correlation. It then compares the embeddings using cosine similarity. Although this is much cheaper than running a neural network for every pair, DeepCoFFEA still performs $O(N^2)$ pairwise comparisons.

Prior correlation methods have not been evaluated at flow counts derived from realistic relay-level or network-level adversaries. We examine this scale in Section~\ref{sec:motivation} and reduce the pairwise-comparison cost using ANN search.

\paragraphB{Approximate Nearest Neighbor Search}
Approximate nearest neighbor (ANN) search retrieves nearby points in a high-dimensional space without exhaustively comparing a query against every indexed point. Common approaches include locality-sensitive hashing (LSH)~\cite{LSH}, inverted-file and product-quantization methods~\cite{IVF}, and graph-based methods such as HNSW~\cite{HNSW} and CAGRA~\cite{CAGRA}. Graph-based ANN methods provide a useful trade-off between search time and recall~\cite{graph_survey}. We use CAGRA in \sys (Section~\ref{sec:attack_w_ann}) and compare it with other ANN indexes in Appendix~\ref{appdneix:ann_comparison}.

\section{Threat Model}
\label{sec:threat_model}
Following prior flow correlation work~\cite{users_get_routed,deepcoffea,deepcorr}, we consider a passive adversary that can observe traffic at both ends of a Tor circuit. We do not consider active attacks that inject, delay, or modify traffic. 
Since Tor traffic is encrypted, the adversary cannot inspect packet contents and instead relies on metadata such as packet timing and size. 
We refer to traffic observed at the entry-guard side as \emph{egress traffic} and traffic observed at the exit-relay side as \emph{ingress traffic}. The adversary's goal is to determine whether an observed egress flow and ingress flow belong to the same communication, thereby linking the client to its destination.

We say that a path is \emph{compromised} when the adversary can observe both ends of the path. The adversary can then apply flow correlation to the two observations in an attempt to identify the matching pair. Thus, observing both ends enables the attack, but does not necessarily imply that the correlation succeeds.


Following prior flow correlation studies such as DeepCorr~\cite{deepcorr} and DeepCoFFEA~\cite{deepcoffea}, we use \emph{flow} to refer to the packet sequence observed at one side of the Tor network for a network connection or session. A Tor circuit can carry multiple streams and may therefore contribute multiple flows over its lifetime.

We consider two types of adversaries, following prior work~\cite{users_get_routed,measuring_mitigating,counter_raptor}: (i) a \emph{relay-level adversary}, which operates Tor relays, and (ii) a \emph{network-level adversary}, which observes Tor traffic from the underlying Internet infrastructure. 

\subsubsection{Relay-level adversary.}
A relay-level adversary operates its own relays in the Tor network and observes traffic that passes through them. To observe both ends of a circuit, the adversary must obtain vantage points at both the guard and exit sides. We characterize the adversary by the bandwidth assigned to its guard and exit relays~\cite{users_get_routed}, since relay-selection probability depends on relay bandwidth.






\subsubsection{Network-level adversary.}
A network-level adversary observes Tor traffic from the Internet paths between clients, relays, and destinations rather than by operating Tor relays. We consider three types of network vantage points: Autonomous Systems (ASes), Internet Exchange Points (IXPs), and IXP organizations. An AS is a collection of IP networks operated under a common routing policy. ASes may be operated by ISPs, enterprises, universities, cloud providers, or other organizations. An IXP is infrastructure through which participating ASes exchange traffic. An IXP organization may operate one or more IXPs and can therefore provide observation points across the IXPs it operates. We consider each AS, IXP, or IXP organization as a separate adversary and do not model collusion among these entities.

\section{Motivation}
\label{sec:motivation}

\subsection{Problems in State-of-the-Art Attacks}

Although DeepCoFFEA~\cite{deepcoffea} reduces the computational cost of DeepCorr~\cite{deepcorr}, it still requires pairwise comparisons between flows. As a result, its computational complexity remains  $O(N^2)$, which limits its scalability as the number of flows increases. In this section, we describe this limitation and how we address it.

DeepCoFFEA~\cite{deepcoffea} uses feature embedding networks (FENs) to map flows into an embedding space where correlated flows have high cosine similarity.  The FENs consist of two separate subnetworks: one for entry-side (egress) traffic and the other for exit-side (ingress) traffic. 
During training, DeepCoFFEA  uses an egress flow as the anchor and pairs with either its corresponding ingress flow (a positive pair) or a non-corresponding ingress flow (a negative pair). A triplet loss is used to bring positive pairs closer in the embedding space while pushing negative pairs farther apart.

Each flow is divided into 11 overlapping time windows, producing 11 embedding vectors per flow. For each window of an egress flow, DeepCoFFEA exhaustively compares it with all $N$ ingress-flow embeddings. This process consists of three steps: computing cosine similarities for all candidate pairs, sorting the similarity scores to select the top-$k$ candidates, and assigning votes to the selected pairs. The votes are then accumulated across windows, and a pair is classified as correlated if its total number of votes exceeds a predefined threshold.

Thus, DeepCoFFEA's exhaustive correlation step includes similarity computation, sorting, and voting. While similarity computation requires $O(WN^2)$, sorting the $N$ candidates for every egress-flow window requires $O(WN^2\log N)$, and voting also operates on the pairwise comparison results.


This approach has two advantages. 
First, it avoids using a neural network as the correlation metric, as done in DeepCorr~\cite{deepcorr}.
DeepCorr requires $N^2$ neural network evaluations over full traces, which results in substantial computation and memory overhead.
Second, DeepCoFFEA operates on relatively short flow segments, unlike approaches such as RAPTOR~\cite{raptor}, which require longer observation periods.

Despite these advantages, DeepCoFFEA still has two important limitations. First, its correlation stage relies on pairwise comparison between egress and ingress flow embedding vectors, resulting in $O(N^2)$ computation. This cost can become prohibitive as the number of observed flows grows.
Second, its original evaluation does not capture the scale and base rate of a realistic attack. We define the \emph{base rate} as the fraction of observed flows that have a corresponding flow on the other side of the network. In practice, an adversary may observe many unrelated flows, and only a fraction of them form true correlated pairs. The base rate therefore directly affects the difficulty of identifying correlated flows, yet the range of base rates that a realistic adversary would encounter has not been well characterized. In this work, we estimate both the number of observable flows and the resulting base rate under different adversary settings.

\begin{figure}[!t]
    \centering
    \includegraphics[width=0.5\linewidth]{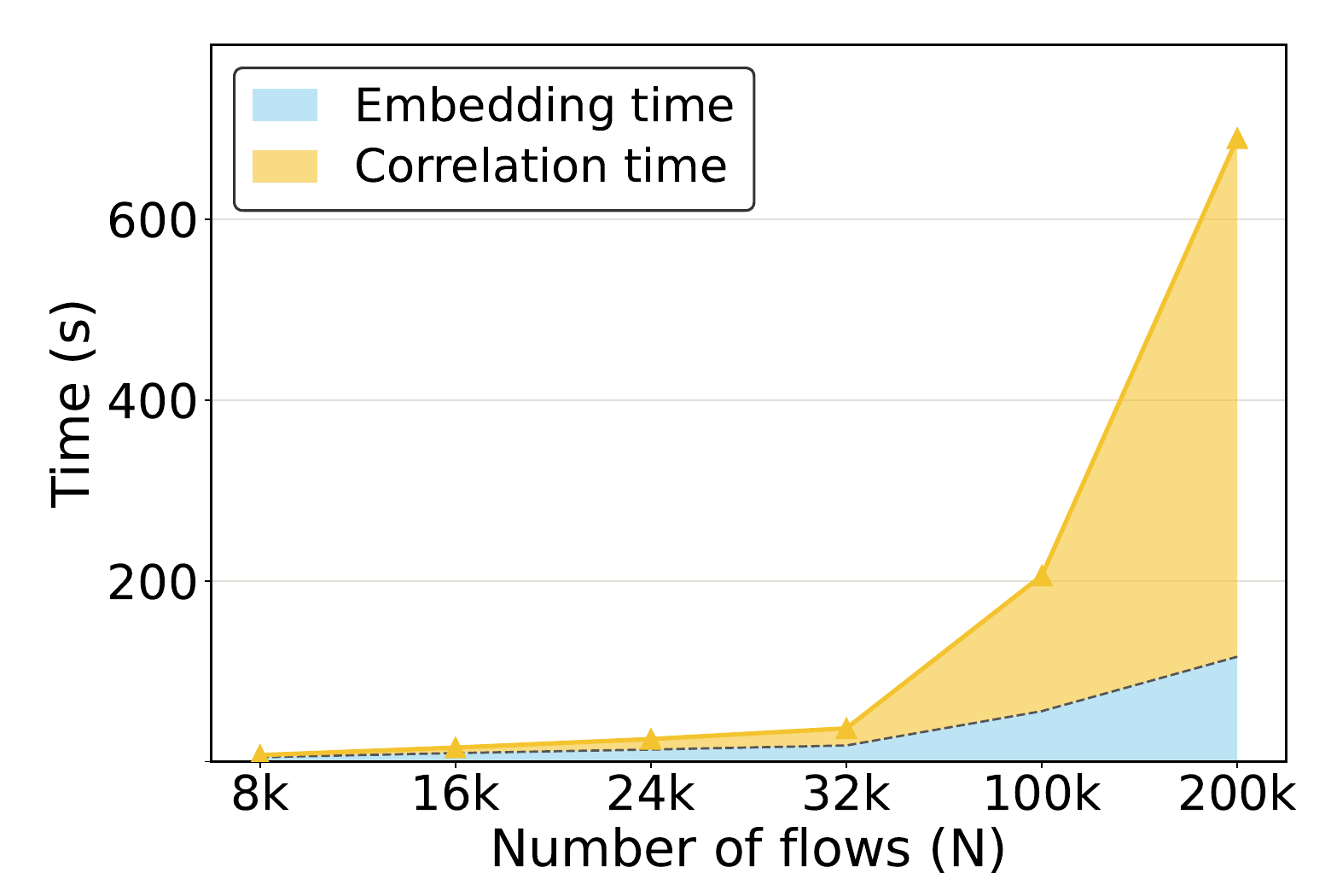}
    \caption{Wall-clock time (seconds) of DeepCoFFEA as the number of flow pairs (N) increases.}
    \label{fig:dcf_scale}
\end{figure}

To examine the first limitation, we measure the wall-clock time of DeepCoFFEA as $N$ increases from 7{,}759 to 199{,}991. 
Figure~\ref{fig:dcf_scale} shows that the runtime increases rapidly with $N$.

For a fair measurement of its computational cost, we first move DeepCoFFEA's correlation operations to the GPU. 
In the original implementation, the similarity matrix multiplication is performed on the GPU, while other operations, including sorting  similarity scores and tallying votes across windows, are performed on the CPU. We implement these operations on the GPU as well to avoid CPU-side overhead that is not inherent to the correlation method itself. 

At the smallest scale, correlation accounts for only 32\% of the total runtime. Its contribution grows quickly with $N$, reaching 52\%  at approximately 32k flows and 83\% at approximately 200k flows. The scaling behavior is also consistent with the quadratic cost of exhaustive comparison. When $N$ doubles from 8k to 16k,  correlation time increases by 2.63$\times$. When $N$ doubles from 100k to 200k, it increases by 3.82$\times$, approaching the 4$\times$ increase expected from quadratic scaling. 

These results show that the computational feasibility of a correlation attack depends on the number of flows an adversary observes. However, $N$ alone does not fully characterize the attack setting: the fraction of those flows that actually have a corresponding flow on the other side, i.e., the base rate, also determines the difficulty of the correlation attack. We therefore characterize both $N$ and the base rate to evaluate correlation attacks under realistic adversary observations.

\subsection{Our Approaches}
\begin{figure}[!t]
    \centering
    \includegraphics[width=0.6\linewidth]{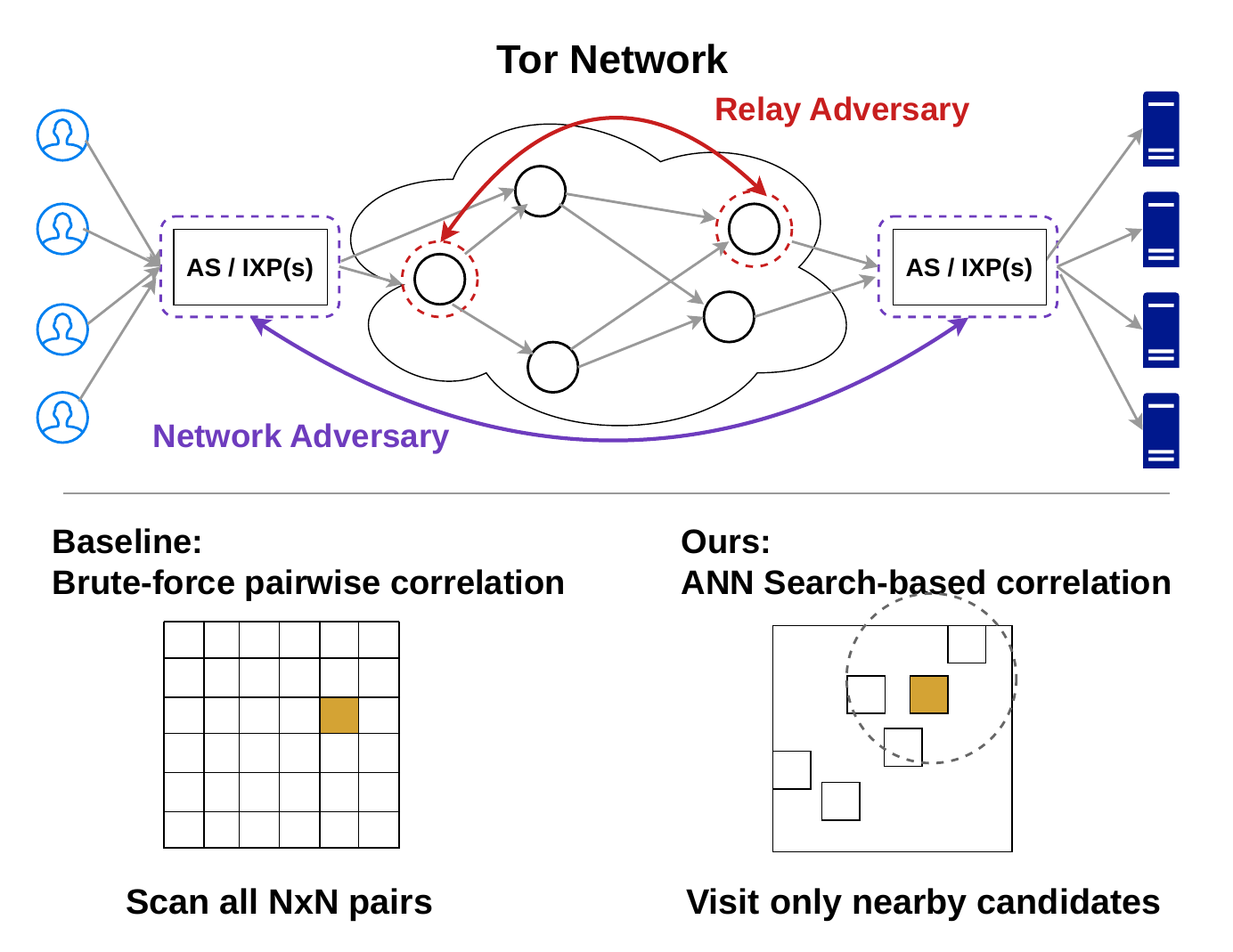}
    \caption{Overview of the threat model and correlation pipeline. Relay- and network-level adversaries observe traffic at both ends of Tor; DeepCoFFEA performs exhaustive pairwise correlation, while InstantCoFFEA uses ANN search to retrieve only nearby candidates.}
    \label{fig:threat_model}
\end{figure}

We address these two limitations in two ways. 
First, we develop a correlation pipeline based on approximate nearest neighbor (ANN) search to reduce the cost of exhaustive pairwise comparison. Figure~\ref{fig:threat_model} gives an overview of our threat model.
Second, we estimate the number of flows that relay-level and network-level adversaries can observe under different network scales and conditions.


DeepCoFFEA identifies candidate matches by comparing each egress flow against all ingress flows. For $N$ egress and $N$ ingress flows, this requires evaluating $N^2$ pairs. Our approach replaces this exhaustive search with ANN search. Given an egress-flow embedding as a query, the ANN index directly retrieves a small set of nearby ingress-flow embeddings without comparing the query against every ingress flow. This removes the need to enumerate all $N^2$ pairs and substantially reduces the cost of candidate retrieval. 

The remaining question is what values of $N$ are realistic for an actual adversary. Estimating this quantity requires accounting for several factors jointly: Tor consensus documents and relay weights, which affect relay-selection probabilities; the location of relays,  clients, and destinations, which determine whether a network-level adversary such as an AS or IXP lies on a circuit path; and the total number of flows generated during the observation period.

These factors are not independent. For example, relay selections across circuit positions are coupled by Tor's path-selection constraints including family and /16 subnet exclusions. Network paths also depend on the particular combinations of clients, relays, and destinations. As a result, it is difficult to estimate the number of flows an adversary observes analytically while accounting for all of these factors together. 

We therefore use simulation to estimate how many flows different adversaries can observe. The simulation uses Tor consensus snapshots, traffic volume, and AS-level topology data to model relay selection and network paths. 


\section{InstantCoFFEA}
\label{sec:attack_w_ann}


In this section, we describe the design of \sys. Up to the embedding stage, \sys follows DeepCoFFEA~\cite{deepcoffea}. Each egress and ingress flow is first divided into 11 overlapping time windows. DeepCoFFEA's two FENs, one for egress traffic and the other for ingress traffic, then map each window to an embedding, producing 11 embeddings per flow.

\sys changes the correlation stage that follows. Instead of comparing every egress embedding with all ingress embeddings, we use CAGRA~\cite{CAGRA} for approximate nearest neighbor (ANN) search. For each window position, we build a CAGRA index over the ingress embeddings and query it with the corresponding egress embeddings to retrieve the top-$k$ candidates. We index the ingress side because it gives better precision-recall performance in our experiments (Appendix~\ref{appendix:indexing side}).

\begin{figure}[h]
    \centering
    \includegraphics[width=0.9\linewidth]{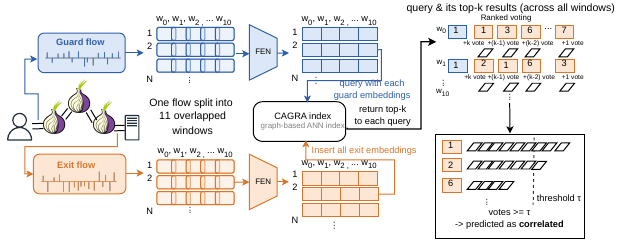}
    \caption{InstantCoFFEA pipeline. Each flow is split into 11 overlapping windows and mapped to embeddings using DeepCoFFEA's FENs. CAGRA indexes the ingress embeddings and retrieves the top-$k$ candidates for each egress-flow window, which are then aggregated using ranked voting.}
    \label{fig:sytsem_overview}
\end{figure}

\subsection{Index Building and Similarity Search}
After obtaining the egress-flow and ingress-flow embeddings, DeepCoFFEA computes cosine similarity for all possible egress-ingress pairs. With $N$ flows on each side, this requires $N^2$ similarity computations for every window.

\sys avoids this exhaustive search by using CAGRA. For each window, CAGRA first builds a proximity graph over the $N$ ingress embeddings. Each embedding is represented as a node, and each node maintains a fixed number $d$ of outgoing edges to nearby nodes.




CAGRA initializes the graph with NN-Descent~\cite{nn-descent} and 
then optimizes the graph by reordering edges and adding reverse edges. 
Edge reordering keeps the more useful neighbors when reducing the graph to degree $d$. For reverse-edge addition, CAGRA reverses the retained edges and merges them back into the graph. This adds paths in the opposite direction and improves reachability between nodes. The reordering and reverse-edge steps require $O(Nd^3)$ and $O(Nd)$, respectively.



Once the index is built, each egress-flow embedding is used as a query. CAGRA maintains a list of the current best candidates and expands the graph from these candidates during search. The search width $p$ controls how many candidates are expanded at each iteration. A larger $p$ can improve search accuracy at the cost of additional computation. The search continues until the candidate list converges or reaches the maximum number of iterations $I_{max}$, after which the $k$ nearest ingress-flow embeddings are returned.

CAGRA does not provide a formal worst-case search bound. For our analysis, we model the search cost as $O(I_{max}pd)$ per query. Since \sys performs one query for each of the $N$ egress flows, the total search cost is $O(NI_{max}pd)$.

\subsection{Candidate Voting}
For each window of an egress flow, CAGRA returns the $k$ nearest ingress-flow embeddings from the corresponding window. Each returned egress-ingress pair is regarded as a candidate match for that window. We then aggregate these candidates across the $W$ windows to determine the final correlated pairs. 

DeepCoFFEA checks all $N^2$ pairs in each window when assigning votes, giving a cost of $O(WN^2)$. After voting, each egress flow can have up to $Wk$ candidates across the $W$ windows. We sort these candidates so that repeated candidates can be grouped and their votes can be accumulated. Sorting $Wk$ candidates takes $O(Wk\log(Wk))$ per egress flow, or $O(NWk\log(Wk))$ over all $N$ egress flows. 

In \sys, CAGRA directly returns the $k$ nearest ingress candidates for each egress-flow window, so no additional sorting is needed to identify the top-$k$ candidates. We then update the vote count for each returned pair across windows. Since there are $O(WNk)$ returned candidates in total, the voting cost is $O(NWk)$.

We consider two voting schemes. In \textit{binary voting}, a candidate pair receives one vote whenever it appears among the top-$k$ neighbors for a window, following the voting scheme of DeepCoFFEA. In \textit{ranked voting}, the vote also reflects the candidate's rank within the returned $k$ neighbors. After all $W$ windows are processed, votes are accumulated for each egress-ingress flow pair. A pair is reported as correlated when its accumulated vote exceeds the threshold $\tau$. 

By using ANN search, \sys reduces both the number of similarity comparisons and the amount of data processed during voting. DeepCoFFEA considers all $N$ ingress candidates for every egress-flow window, whereas \sys keeps only the $k$ candidates returned by CAGRA. Table~\ref{tab:time_complexity} summarizes the computational cost of each stage.

\begin{table}[!t]
\centering
\fontsize{8}{9}\selectfont
\renewcommand{\arraystretch}{0.95}
\setlength{\tabcolsep}{3pt}
\caption{Per-stage time complexity of DeepCoFFEA and InstantCoFFEA. Here, $W=11$, and $d$, $k$, $p$, and $I_{\max}$ are treated as constants.}
\label{tab:time_complexity}
\begin{tabular}{@{}lc@{\hspace{10pt}}lc@{}}
\toprule
\multicolumn{2}{c}{\textbf{DeepCoFFEA}} & \multicolumn{2}{c}{\textbf{\sys}} \\
\cmidrule(r){1-2}\cmidrule(l){3-4}
Embedding  & $O(WN)$                         & Embedding   & $O(WN)$ \\
Sim. Comp. & $O(WN^2)$                       & Index Build & $O(WNd^2\log_d N + WNd^3 + WNd)$ \\
Sorting    & $O(WN^2\log N)$                 & Searching   & $O(WNI_{max}pd)$ \\
Voting     & $O(N^2W + NWk\log Wk)$         & Voting      & $O(NWk)$ \\
\midrule
\textbf{Total} & $\boldsymbol{O(N^2\log N)}$ & \textbf{Total} & $\boldsymbol{O(N\log N)}$ \\
\bottomrule
\end{tabular}
\end{table}

\subsection{Computational Cost Analysis}

Table~\ref{tab:time_complexity} summarizes the time complexity of DeepCoFFEA and \sys. With $W$, $d$, $k$, $p$, and $I_{\max}$ treated as constants, DeepCoFFEA has an overall complexity of $O(N^2\log N)$, while \sys has $O(N\log N)$. However, the constant of $d$ in the CAGRA index construction (i.e., $O(Nd^3)$) can still be significant, especially for smaller $N$. We therefore substitute the parameter values used in our experiments and compare the resulting operation counts.


We set $W=11$, following DeepCoFFEA's window configuration. 
For the remaining parameters, we conduct a parameter sweep as described in Appendix~\ref{appendix:hyperpaarmeter}. 
For the CAGRA search parameters, we select $p=1$, $I_{max}=139$ based on the accuracy-latency trade-off. We set the graph degree to $d=8$, as detection performance shows little improvement beyond this value. For the number of returned neighbors, we sweep $k \in \{2, 4, \dots 1024 \}$ and obtained the highest area under precision-recall curve (PR-AUC) at $k=32$.

Using these values, we compute the operation counts of DeepCoFFEA and InstantCoFFEA from the complexity functions in Table~\ref{tab:time_complexity}. The detailed calculation is provided in Appendix~\ref{appendix:extended_time_complexity}. The estimated operation count of DeepCoFFEA becomes larger than that of \sys when $N>2$. This point is well below the scale used in DeepCoFFEA's original evaluation, indicating that the additional cost of CAGRA index construction is offset even at a relatively small number of flows.

\section{Evaluation}
\label{sec:evaluation}
We evaluate \sys with respect to the two limitations discussed in Section~\ref{sec:motivation}. Our evaluation is organized around the following research questions:
\begin{itemize}
    \item[\textbf{RQ1}] \textbf{Scalability and detection performance:} How much does the ANN-based pipeline reduce wall-clock time, and how does it affect detection performance compared with prior work?
    \item[\textbf{RQ2}] \textbf{Performance under realistic conditions:} How do \sys and prior approaches perform under realistic flow counts and base rates?
\end{itemize}

\subsection{Dataset and Experimental Setup}
We use the dataset released by DeepCoFFEA~\cite{deepcoffea}, collected by crawling 60,084 Alexa websites and containing 60,084 unique flow pairs. We split the dataset into training and test sets, and train the FENs for 1,000 epochs on an NVIDIA RTX A6000 GPU with 48GB of memory. 

For detection performance, we consider two settings. The \emph{lab setting} follows prior work, where $N=N_{guard}=N_{exit}$ and every observed flow has a matching flow on the other side. The \emph{realistic setting} uses $N=(N_{guard},N_{exit})$ in which $N_{guard}\neq N_{exit}$, and the base rate $\pi = N_{\mathrm{match}}/(N_{\mathrm{guard}}\cdot N_{\mathrm{exit}})$, defined as the fraction of candidate pairs that are true matching pairs, is obtained from our simulation.
To measure wall-clock time beyond the size of the original dataset, we increase $N$ by duplicating flows and adding timing jitter. This is sufficient for runtime measurement because the computation cost mainly depends on the number of input flows.\footnote{Appendix~\ref{appendix:jitter_vs._real} compares runtime using real flows and duplicated \& jittered flows. The duplicated and jittered data requires approximately 93.5\% of the wall-clock time measured on real data.}

\subsection{Scalability and Detection Performance}

\subsubsection{Runtime and Scalability.}
To answer RQ1, we measure the wall-clock time as $N$ increases from 7{,}759 to 199{,}991 (approximately 8k to 200k). Each configuration is run three times, and we report the mean; the standard deviation is below 5\% for all configurations.

\begin{figure}[!t]
    \centering
    \includegraphics[width=\linewidth]{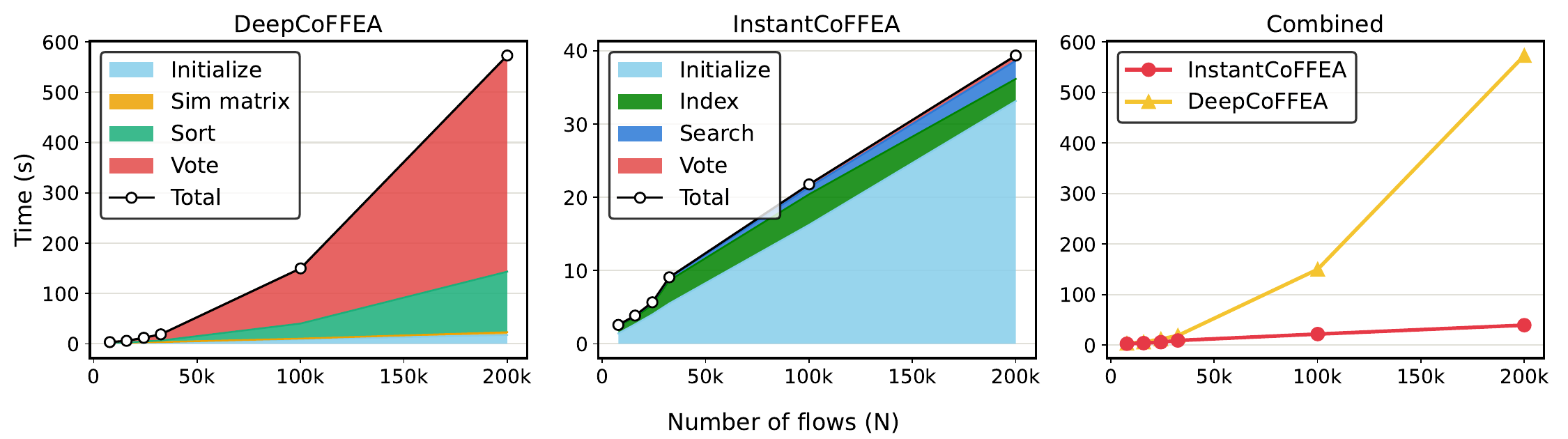}
    \caption{Wall-clock time (seconds) breakdown of DeepCoFFEA (left) and InstantCoFFEA (middle) as $N$ increases; the right panel compares their total runtimes.}
    \label{fig:op_time_breakdown}
\end{figure}

Figure~\ref{fig:op_time_breakdown} shows the runtime breakdown of DeepCoFFEA and \sys. Embedding computation is common to both methods and is therefore excluded from Figure~\ref{fig:op_time_breakdown}.\footnote{The embedding times are 4.89\,s, 9.44\,s, 13.34\,s, 18.84\,s, 55.85\,s, and 116.88\,s for $N$ = 7,759, 15,866, 24,210, 32,441, 100{,}058, and 199,991, respectively.} Excluding embedding, DeepCoFFEA's runtime increases from 3.35\,s at $N$ = 7,759 to 572.94\,s at $N$=199,991. Over the same range, \sys increases from 2.64\,s to 39.35\,s.

For DeepCoFFEA, similarity computation itself remains relatively small, increasing from 0.01\,s to 5.43\,s because the similarity matrix is computed using GPU matrix multiplication. The larger costs come from sorting, which increases from 0.18\,s to 120.38\,s, and voting, which increases from 0.95\,s to 429.55\,s. At $N$=199,991, sorting and voting together account for 96.0\% of the total runtime excluding embedding computation time. 

\sys avoids these two costs over the full $N^2$ similarity matrix. CAGRA index construction increases from 1.03\,s to 3.02\,s, while search increases from 0.10\,s to 2.50\,s. Since CAGRA directly returns the top-$k$ candidates, with $k=32$ in these experiments, no separate sorting over all ingress flows is needed. Voting is performed only on the returned candidates, taking from 0.03\,s to 0.70\,s.

We also report the correlation time per flow pair. At $N=100{,}058$, \sys requires 0.22\,ms, compared with 1.50\,ms for DeepCoFFEA, giving a $6.89 \times$ speedup. At $N=199{,}991$, the corresponding times are 0.20\,ms and 2.86\,ms, for a $14.56\times$ speedup. The gap increases with $N$, consistent with the different scaling behavior shown in Table~\ref{tab:time_complexity}.


\subsubsection{Detection Performance.}
\begin{figure}[!t]
    \centering
    \begin{subfigure}[b]{0.4\textwidth}
       \includegraphics[width=\linewidth]{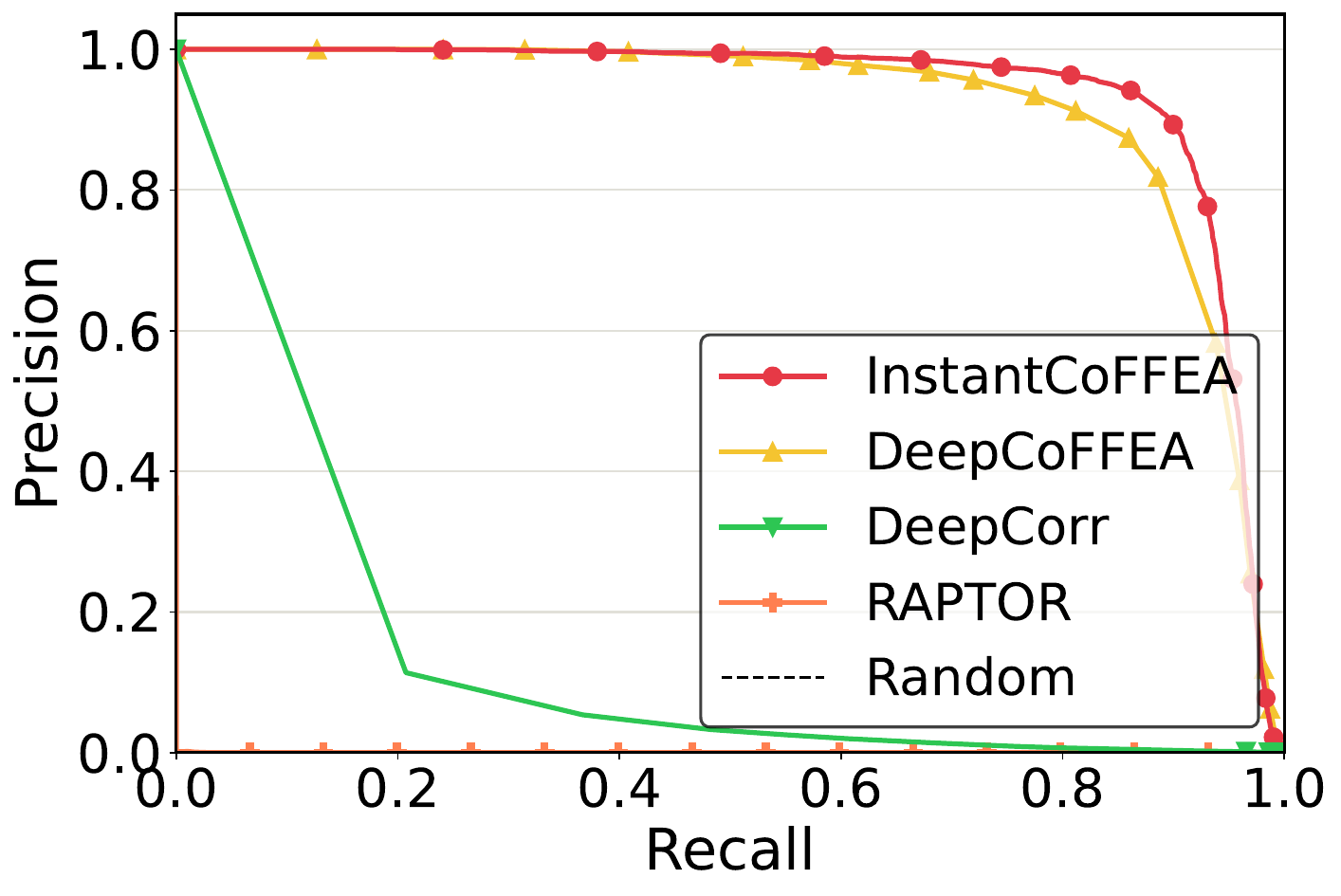}
       \caption{Lab setting}
       \label{subfig:pr_lab}
    \end{subfigure}
     \begin{subfigure}[b]{0.4\textwidth}
       \includegraphics[width=\linewidth]{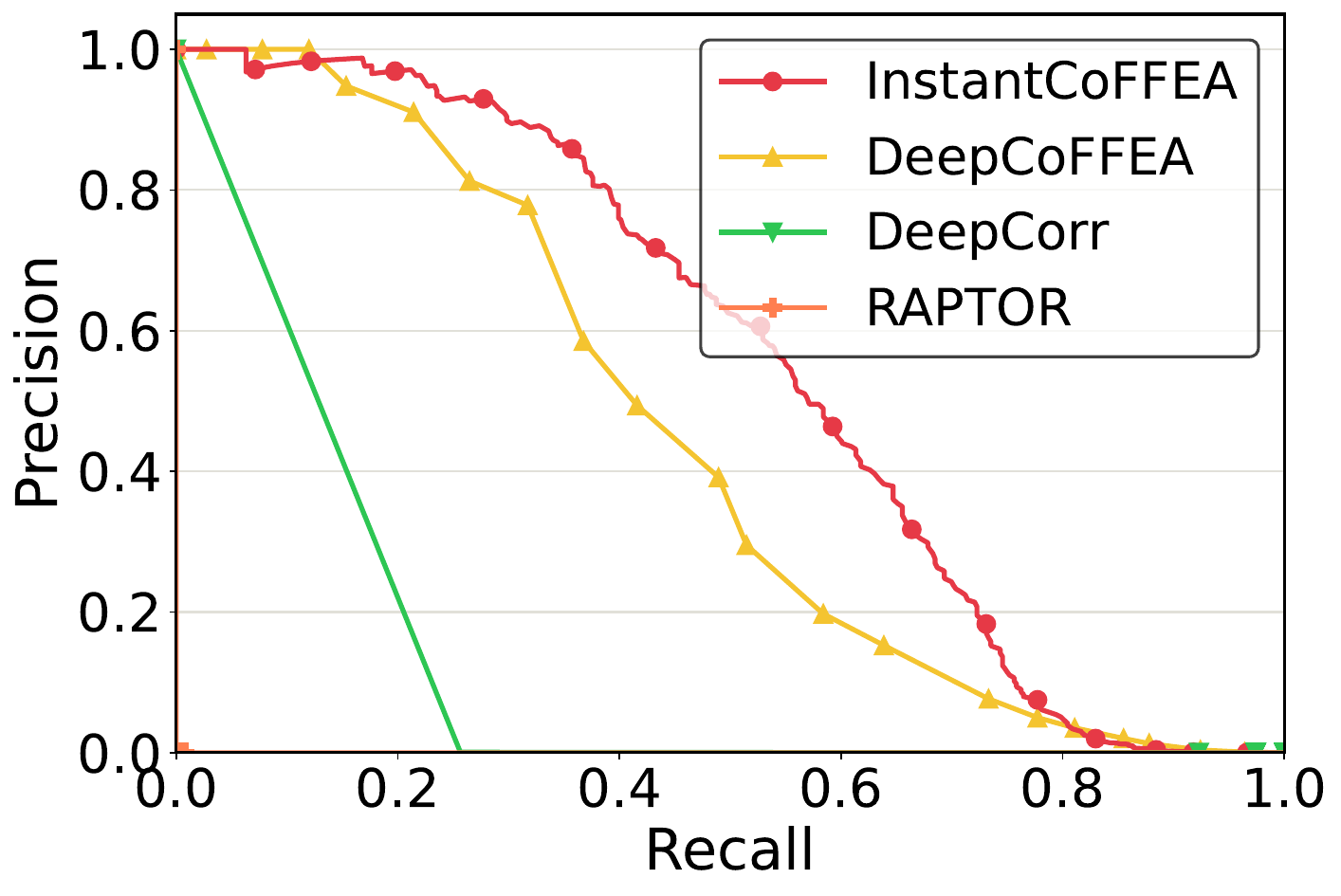}
       \caption{Realistic setting}
       \label{subfig:pr_real}
    \end{subfigure}
    \caption{Precision--recall curves of InstantCoFFEA and prior methods under the lab setting (left) and realistic setting (right)}
    \label{fig:pr_curve}
\end{figure}

We next evaluate detection performance in the lab setting using the precision--recall curves. Figure~\ref{subfig:pr_lab} compares \sys with DeepCoFFEA~\cite{deepcoffea}, DeepCorr~\cite{deepcorr}, and RAPTOR~\cite{raptor}. For \sys, we repeat the evaluation over five random test splits\footnote{Appendix~\ref{appenix:prauc_as_N_rows} reports PR-AUC as $N$ increases. \sys consistently achieves higher PR-AUC than DeepCoFFEA across various $N$s.}. 
Despite using approximate nearest-neighbor search, \sys does not lose detection performance. Its mean PR-AUC is 0.9621, compared with 0.9433 for DeepCoFFEA. Across the five test splits, the standard deviations are 0.0033 and 0.0021, respectively.

We examine two reasons for this result. First, CAGRA retrieves nearly the same top-$k$ candidates as exhaustive search. We measure recall@$k$ against brute-force search for each sub-flow while varying $k$, and observe only a small difference between the two (Table~\ref{tab:recall_at_k_dcf_icf} in Appendix~\ref{appendix:top_k_count}). 


Second, \sys uses the rank of each returned candidate during voting. DeepCoFFEA uses binary voting: a pair receives one vote when it appears among the top-$k$ candidates for a window. \sys supports the same binary voting scheme, but we also evaluate \emph{ranked voting}, which gives larger weights to candidates appearing higher in the returned list (i.e., better candidates). Ranked voting gives better detection performance than binary voting; the full comparison is provided in Appendix~\ref{appendix:voting_scheme}.

Appendix~\ref{appendix:top_k_count} further reports the rank distribution of true pairs and the number of true pairs that remain above the final correlation threshold. Ranked voting retains more true pairs above the threshold than binary voting. This advantage remains at low false-positive rates, including FPRs down to $10^{-8}$.

\subsection{Performance under Realistic Attack Conditions}

\subsubsection{Simulation Setup.}
To estimate the flow counts and base rates seen by an adversary, we simulate the Tor network using TorPS~\cite{users_get_routed} with the May 15, 2026 consensus~\cite{collecTor}. For each 10-minute observation window, we generate circuits across the network, assign destinations and flow types, and trace the relays, ASes, and IXPs along each path. Following prior work~\cite{deepcorr,deepcoffea}, we treat each observed circuit as one flow. 

We repeat the simulation for 24 hourly consensus  snapshots over a 24-hour period. For each snapshot, we record the number of flows observed on the entry-guard side ($N_{\mathrm{guard}}$) and exit side ($N_{\mathrm{exit}}$), as well as the number of matching flows ($N_{\mathrm{match}}$). We define the compromise rate as $R_{\mathrm{comp}}=N_{\mathrm{match}}/N_{\mathrm{total}}$, and the base rate as $\pi=N_{\mathrm{match}}/(N_{\mathrm{guard}}\cdot N_{\mathrm{exit}})$, i.e., the fraction of candidate flow pairs that are true matches. 

We update TorPS's path-selection algorithm to reflect the current Tor implementation with Tor version 0.4.9, including family identifiers, the current guard-selection algorithm, and IPv6 /32 subnet exclusion.
\footnote{We do not model Conflux (Tor 0.4.8), as no existing Tor correlation dataset captures Conflux traffic; we leave its analysis to future work.
}


For destinations, we consider both per-country and global settings. Nithyanand \textit{et al.}~\cite{measuring_mitigating} use country-specific censored-site lists from Citizen Lab~\cite{citizen_lab}, while Barton \textit{et al.}~\cite{DeNASA} sample destinations from the Alexa top 200~\cite{alexa}. Since Alexa is no longer available, we use the Tranco list~\cite{tranco_list} and sample destinations according to a Zipf distribution with $s=1.0$, following the site-visit distribution reported by Adamic and Huberman~\cite{zipf}.

For the relay adversary, we consider total bandwidths of $2{,}000$ and $4{,}000$\,MB/s, corresponding to $1.05\%$ and $2.10\%$ of the total Tor network bandwidth. Both are below the $2.92\%$ adversarial bandwidth considered by Johnson \textit{et al.}~\cite{users_get_routed}.

For network-level adversaries, we build the AS topology using RouteViews BGP tables~\cite{routeviews}, the CAIDA ITDK dataset~\cite{CAIDA_ITDK}, and the CAIDA AS-relationship dataset~\cite{CAIDA_as_relationship}. We infer AS paths using the Qiu and Gao algorithm~\cite{qiu}, and obtain IXP and organization information from the CAIDA IXP~\cite{CAIDA_IXP} and AS-to-organization~\cite{CAIDA_AS2ORG} datasets.




\begin{table}[!t]
\centering
\caption{Simulated guard- and exit-side flow counts and compromise rates for relay- and network-level adversaries.}
\label{tab:scale_top1}
\resizebox{0.99\textwidth}{!}{%
\begin{tabular}{llccccc}
\toprule
Adversary settings &  & 1.4M & 3M & 6M & 9M & Compromise rate \\
\midrule
\multirow{2}{*}{2,000 MB/s (relay)} & Guard & 13,889 $\pm$ 329 & 36,039 $\pm$ 843 & 60,014 $\pm$ 1,170 & 89,247 $\pm$ 1,731 & \multirow{2}{*}{0.0240\% $\pm$ 0.0019\%} \\
 & Exit & 36,300 $\pm$ 1,480 & 93,117 $\pm$ 3,668 & 142,488 $\pm$ 4,637 & 213,833 $\pm$ 7,068 & \\
\midrule
\multirow{2}{*}{4,000 MB/s (relay)} & Guard & 27,386 $\pm$ 647 & 70,246 $\pm$ 1,298 & 117,652 $\pm$ 2,156 & 177,226 $\pm$ 3,390 & \multirow{2}{*}{0.0903\% $\pm$ 0.0058\%} \\
 & Exit & 70,438 $\pm$ 2,637 & 167,063 $\pm$ 5,387 & 278,506 $\pm$ 8,644 & 417,457 $\pm$ 12,948 & \\
\midrule
\multirow{2}{*}{AS174 (AS)} & Guard & 255,968 $\pm$ 6,082 & 547,391 $\pm$ 13,052 & 1,094,497 $\pm$ 25,882 & 1,639,872 $\pm$ 38,554 & \multirow{2}{*}{2.9009\% $\pm$ 0.0459\%} \\
 & Exit & 186,358 $\pm$ 3,756 & 399,120 $\pm$ 7,719 & 798,242 $\pm$ 15,837 & 1,195,155 $\pm$ 23,785 & \\
\midrule
\multirow{2}{*}{Amsterdam IX (IXP)} & Guard & 282,661 $\pm$ 6,448 & 606,431 $\pm$ 13,743 & 1,210,416 $\pm$ 27,511 & 1,814,440 $\pm$ 40,992 & \multirow{2}{*}{12.9146\% $\pm$ 0.0885\%} \\
 & Exit & 657,497 $\pm$ 19,472 & 1,409,261 $\pm$ 42,312 & 2,818,273 $\pm$ 84,131 & 4,222,067 $\pm$ 125,959 & \\
\midrule
\multirow{2}{*}{Amsterdam IX B.V. (IXP Org)} & Guard & 282,726 $\pm$ 6,449 & 606,552 $\pm$ 13,747 & 1,210,688 $\pm$ 27,518 & 1,814,844 $\pm$ 41,002 & \multirow{2}{*}{12.9176\% $\pm$ 0.0885\%} \\
 & Exit & 657,503 $\pm$ 19,472 & 1,409,273 $\pm$ 42,312 & 2,818,299 $\pm$ 84,131 & 4,222,105 $\pm$ 125,960 & \\
\bottomrule
\end{tabular}%
}
\end{table}

\subsubsection{Observed Flow Counts and Base Rates.}
For the relay adversary, we consider two total bandwidths (2{,}000 and 4{,}000MB/s) and four guard-to-exit bandwidth ratios (1:1, 3:1, 6:1, and 1:3). For the network adversary, we consider the two destination settings described above. In both cases, we vary the total number of circuits in the network from 1.4M to 9M. 
Table~\ref{tab:scale_top1} summarizes the relay configuration with the highest compromise rate and, for each network-adversary type, only the top-ranked entity by mean compromise rate. The top-five network-adversary results and additional destination settings are reported in Appendix~\ref{appendix:network_adv}.

For the relay adversary, the 1:3 guard-to-exit bandwidth ratio used in Table~\ref{tab:scale_top1} yields the highest compromise rate among the configurations we tested. Increasing the total number of circuits from 1.4M to 9M has little effect on the compromise rate. Doubling the adversarial bandwidth, however, roughly  doubles the number of observed flows and increases the  compromise rate by about 3.7$\times$.

For network-level adversaries, the number of observed flows varies substantially across entities. The relationship between $N_{\mathrm{guard}}$ and $N_{\mathrm{exit}}$ also differs across AS, IXP, and IXP-organization adversaries. Among the top-ranked entities shown in Table~\ref{tab:scale_top1}, AS174 has the highest mean compromise rate for AS adversaries, while Amsterdam IX ranks first for IXPs. The top IXP organization is also Amsterdam IX because that organization consists of a single large IXP. Lower-ranked IXP organizations can include multiple IXPs, so their results do not necessarily match those of any single IXP.

\begin{figure}[!t]
    \centering
    \includegraphics[width=\linewidth]{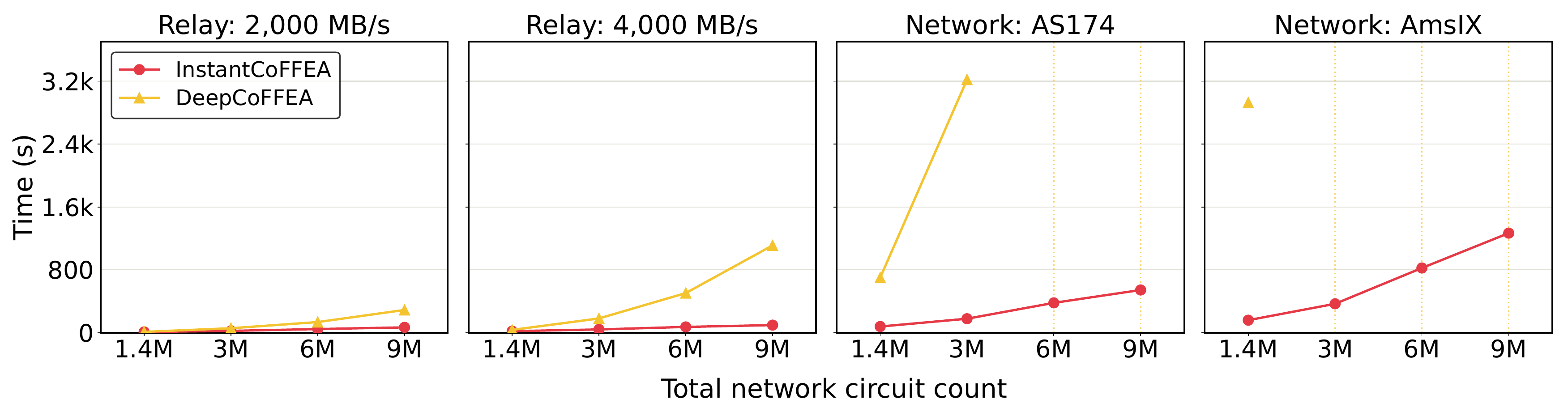}
    \caption{Wall-clock time of InstantCoFFEA and DeepCoFFEA under realistic relay- and network-level adversary settings. Missing DeepCoFFEA points indicate out-of-memory (OOM) failures.}
    \label{fig:op_time_real}
\end{figure}

\subsubsection{Runtime at Realistic Scale.}
We next use the simulated $N_{\mathrm{guard}}$ and $N_{\mathrm{exit}}$ values to measure runtime at the scales seen by these adversaries. Figure~\ref{fig:op_time_real} shows the wall-clock time of DeepCoFFEA and \sys. As in Figure~\ref{fig:op_time_breakdown}, embedding time is excluded because it is common to both methods. \footnote{Embedding time ranges from 27.17\,s  for the 2{,}000 MB/s relay adversary at the 1.4M-circuit scale to 3,766.1\,s for Amsterdam IX at the 9M-circuit scale.} 


For the 2{,}000~MB/s relay adversary, \sys takes 11.04\,s to 69.08\,s as the network scale increases, compared with 11.94\,s to 290.36\,s for DeepCoFFEA. At 4{,}000~MB/s, \sys takes 20.51\,s to 98.00\,s, while DeepCoFFEA takes 36.95\,s to 1,111.48\,s. 

The difference becomes larger for network-level adversaries. DeepCoFFEA runs out of memory when constructing the $N_{guard} \times N_{exit}$ similarity matrix for AS174 at the 6M and 9M scales, and for Amsterdam IX at the 3M--9M scales. \sys does not construct this full matrix and remains runnable at these scales. 
For AS174, DeepCoFFEA's GPU memory reaches 27.38\,GB at the 3M scale and exceeds the available 48GB at the next scale (6M). For Amsterdam IX, it already uses 44.05\,GB at the 1.4M scale and likewise runs out of memory at the next scale (3M).
\sys, in contrast, uses only 0.58--1.81\,GB of GPU memory for AS174 and 0.87--3.25\,GB for Amsterdam IX across all scales tested, leaving ample headroom on the GPU.
Its runtime ranges from 79.95\,s to 544.51\,s for AS174 and from 160.47\,s to 1,267.52\,s for Amsterdam IX. 
The full memory usage of both methods is in Appendix~\ref{appendix:memory_spec}.

\subsubsection{Detection Performance at Realistic Base Rates.}
Finally, we evaluate detection performance using the flow counts and base rates obtained from the simulation. We use the 2{,}000~MB/s relay-adversary setting with $N_{\mathrm{guard}}=13{,}889$ and $N_{\mathrm{exit}}=36{,}300$. Among the candidate pairs, only 352 are true matches, giving a base rate of  $\pi = 6.98\times10^{-5}\%$. Figure~\ref{subfig:pr_real} compares \sys with DeepCoFFEA, DeepCorr, and RAPTOR under this setting. All methods perform worse than in the lab setting because true matches make up a much smaller fraction of the candidate pairs. False positives therefore have a larger effect on precision.


The performance gap between \sys and DeepCoFFEA also increases in this setting. \sys achieves a mean PR-AUC of 0.5335, compared with 0.4172 for DeepCoFFEA. The standard deviations across the five test splits are 0.0278 and 0.0250, respectively. Ranked voting is particularly useful at this low base rate because candidates that repeatedly appear near the top of the returned neighbor list receive more weight than lower-ranked candidates.

\section{Conclusion}
We present InstantCoFFEA, which combines ANN search with ranked voting to improve both runtime and precision-recall over DeepCoFFEA. We also estimate realistic flow counts and base rates through simulation, showing that prior evaluations do not capture the sparsity of true matching pairs. Our simulation has one limitation: TorPS models circuits rather than traffic, and we follow prior work by treating each observed circuit as one flow. Since a circuit may carry multiple streams, actual flow counts may be higher. Overall, our results suggest that computation is no longer the main obstacle to large-scale correlation; the low base rate is.

\bibliographystyle{splncs04}
\bibliography{references}

\appendix
\section{ANN comparison}
\label{appdneix:ann_comparison}

\begin{figure}[H]
    \centering
    \begin{subfigure}[b]{0.49\textwidth}
        \includegraphics[width=\linewidth]{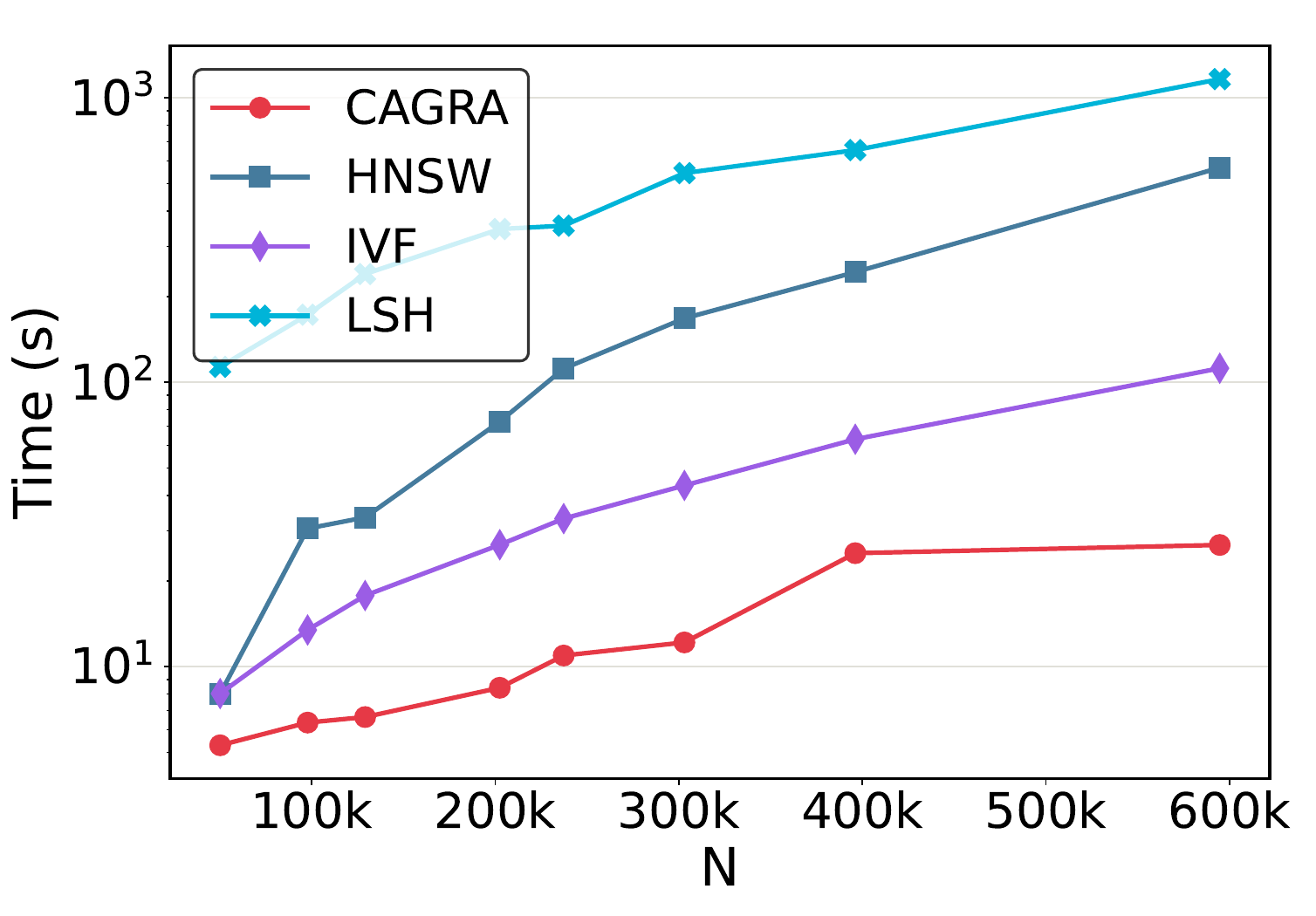}
    \caption{Correlation time as $N$ grows}
    \end{subfigure}
    \begin{subfigure}[b]{0.49\textwidth}
        \includegraphics[width=\linewidth]{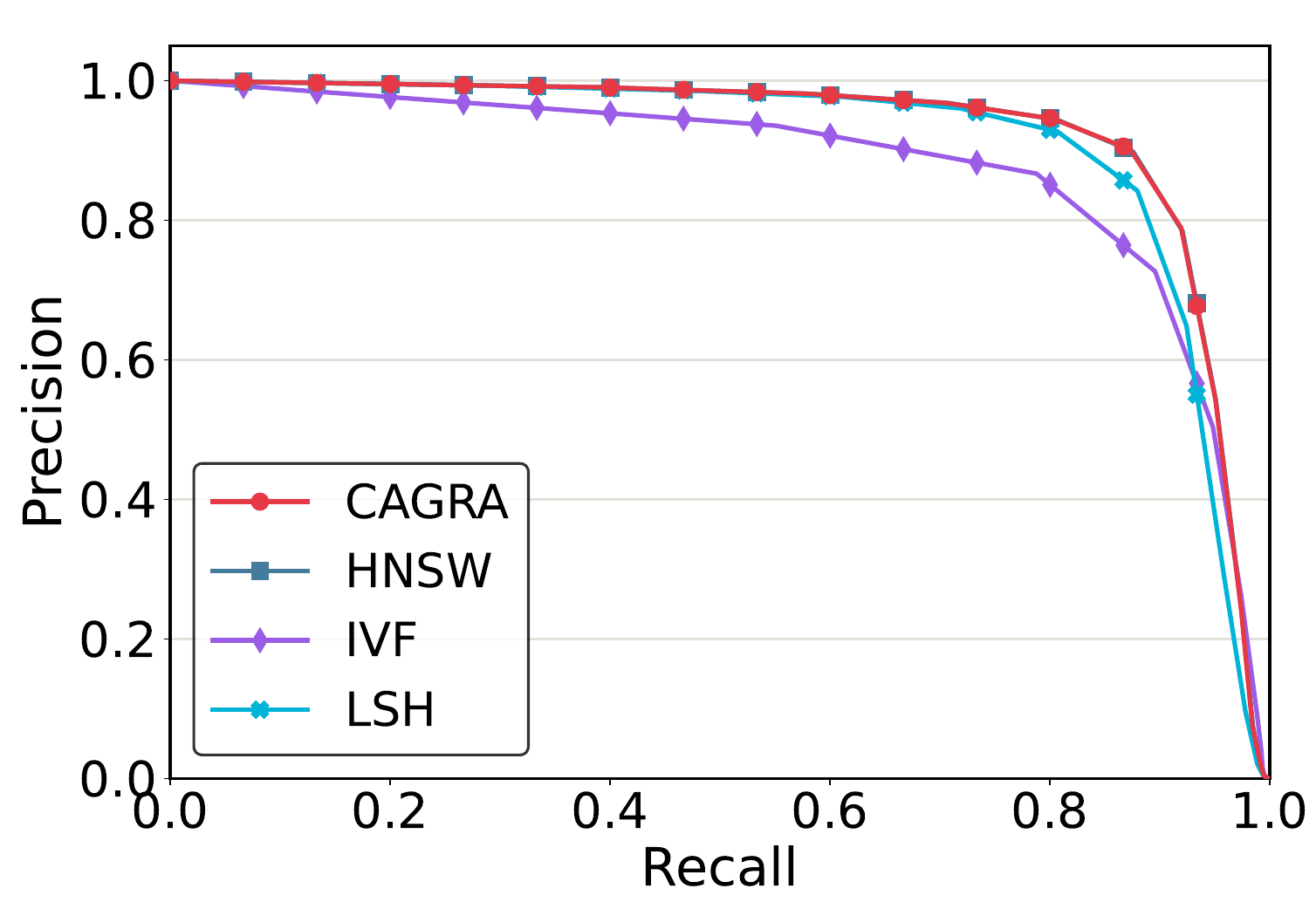}
    \caption{Precision-Recall Curve}
    \end{subfigure}
    \caption{ANN methods comparison in correlation time and precision-recall curve}
    \label{fig:ann_comparison}
\end{figure}

Figure~\ref{fig:ann_comparison} compares four ANN methods: CAGRA, HNSW, IVF, and LSH in correlation time and precision-recall. CAGRA achieves the best recall at a given operating time, which motivated its adoption in InstantCoFFEA.

\section{Indexing Side}
\label{appendix:indexing side}
\begin{figure}[H]
    \centering
    \begin{subfigure}[b]{0.49\textwidth}
        \includegraphics[width=\linewidth]{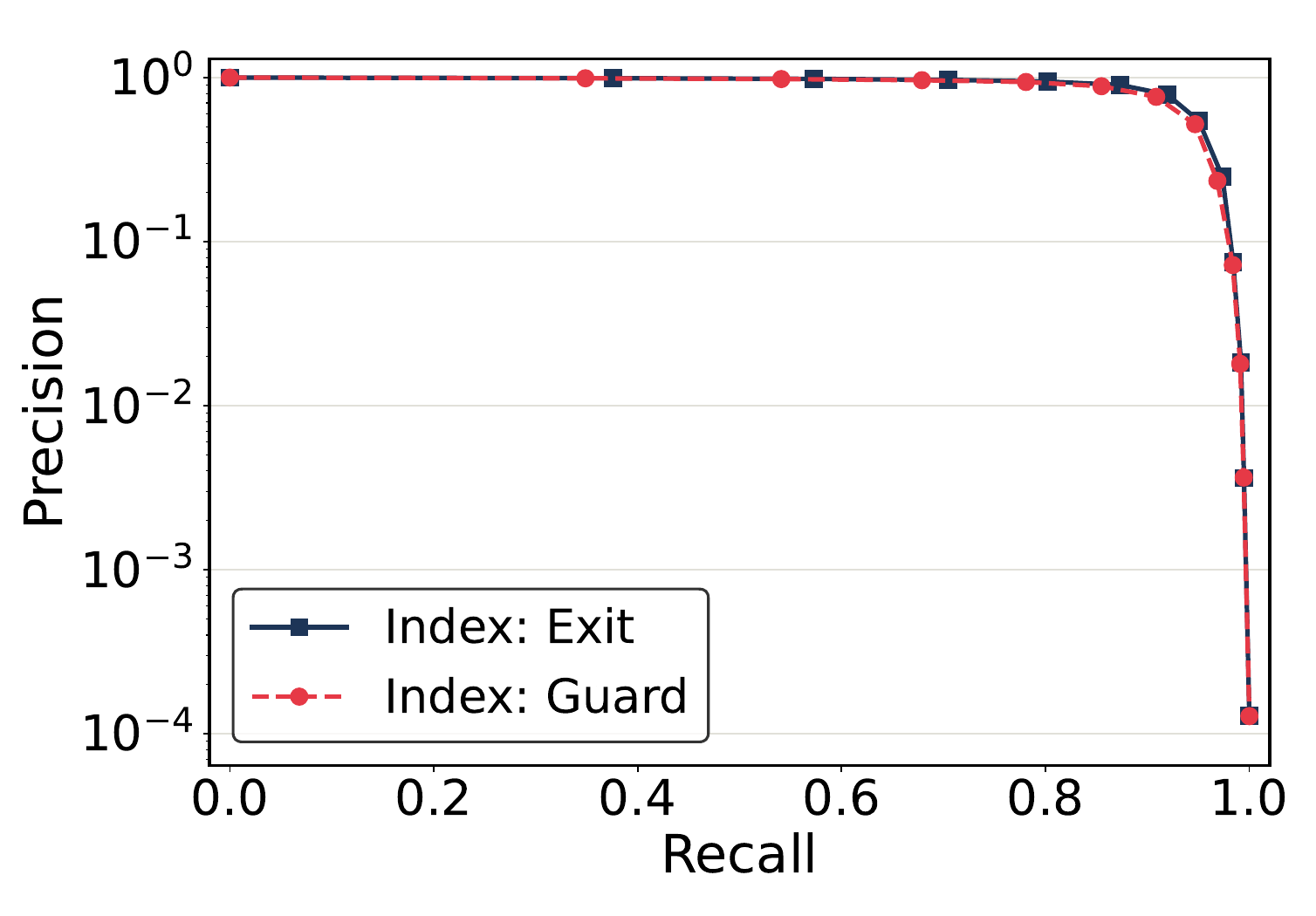}
        \caption{Lab setting}
        \label{subfig:pr_cagra_swap_lab}
    \end{subfigure}
    \begin{subfigure}[b]{0.49\textwidth}
        \includegraphics[width=\linewidth]{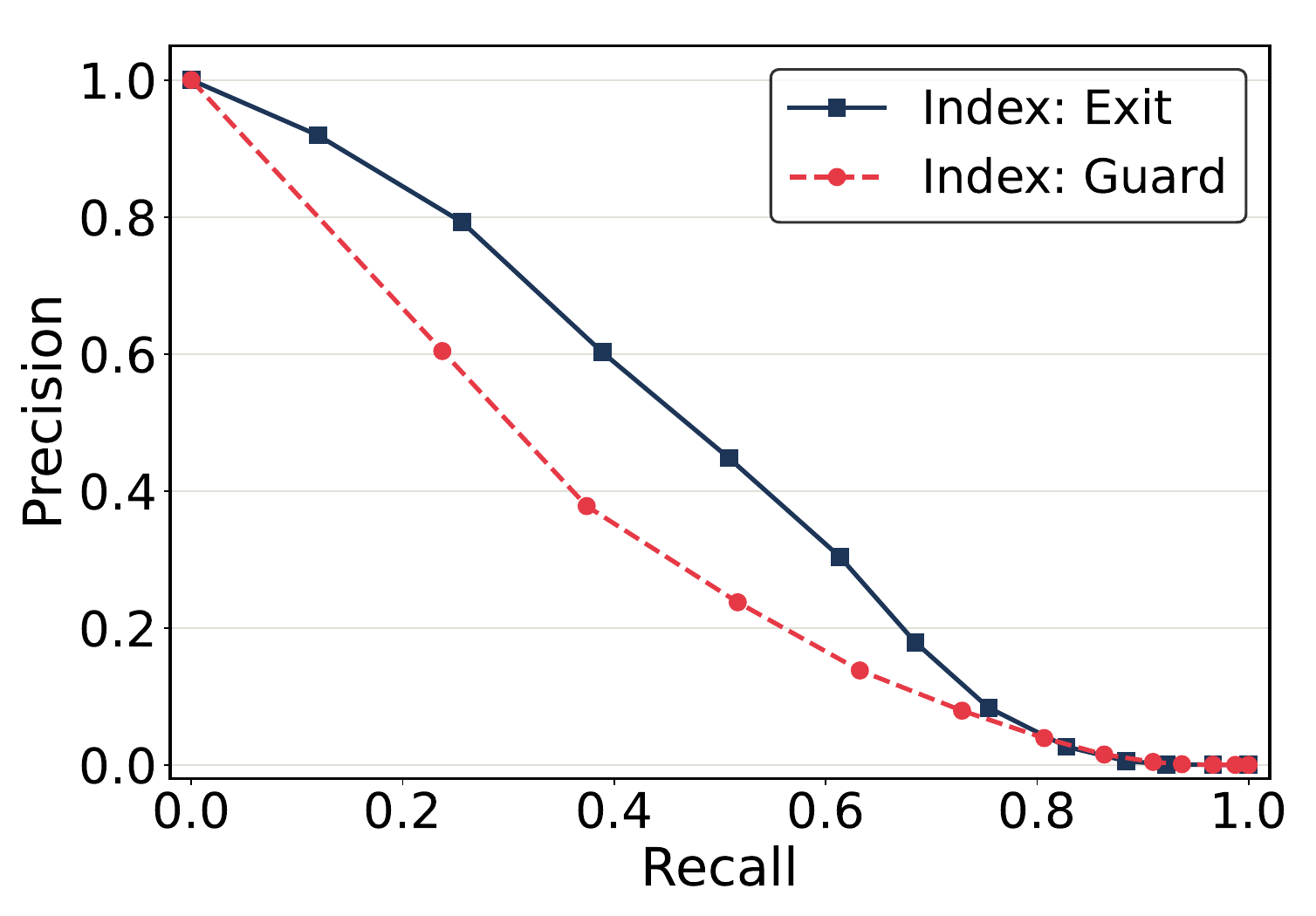}
        \caption{Realistic setting}
        \label{subfig:pr_cagra_swap_real}
    \end{subfigure}
    \caption{PR Curve of exit-side vs. guard-side indexing}
    \label{fig:PROC_cagra_swap}
\end{figure}

Figure~\ref{fig:PROC_cagra_swap} compares precision-recall when indexing
exit-side vs. guard-side embeddings under both the lab and realistic settings. 
The two indexing directions perform comparably in the lab setting, but exit-side indexing achieves clearly higher precision-recall in the realistic setting, motivating our design choice. 

\section{CAGRA Hyperparameter Tuning}
\label{appendix:hyperpaarmeter}

\subsection{Graph degrees}

\begin{figure}[H]
    \centering
    \begin{subfigure}[b]{0.49\textwidth}
        \includegraphics[width=\linewidth]{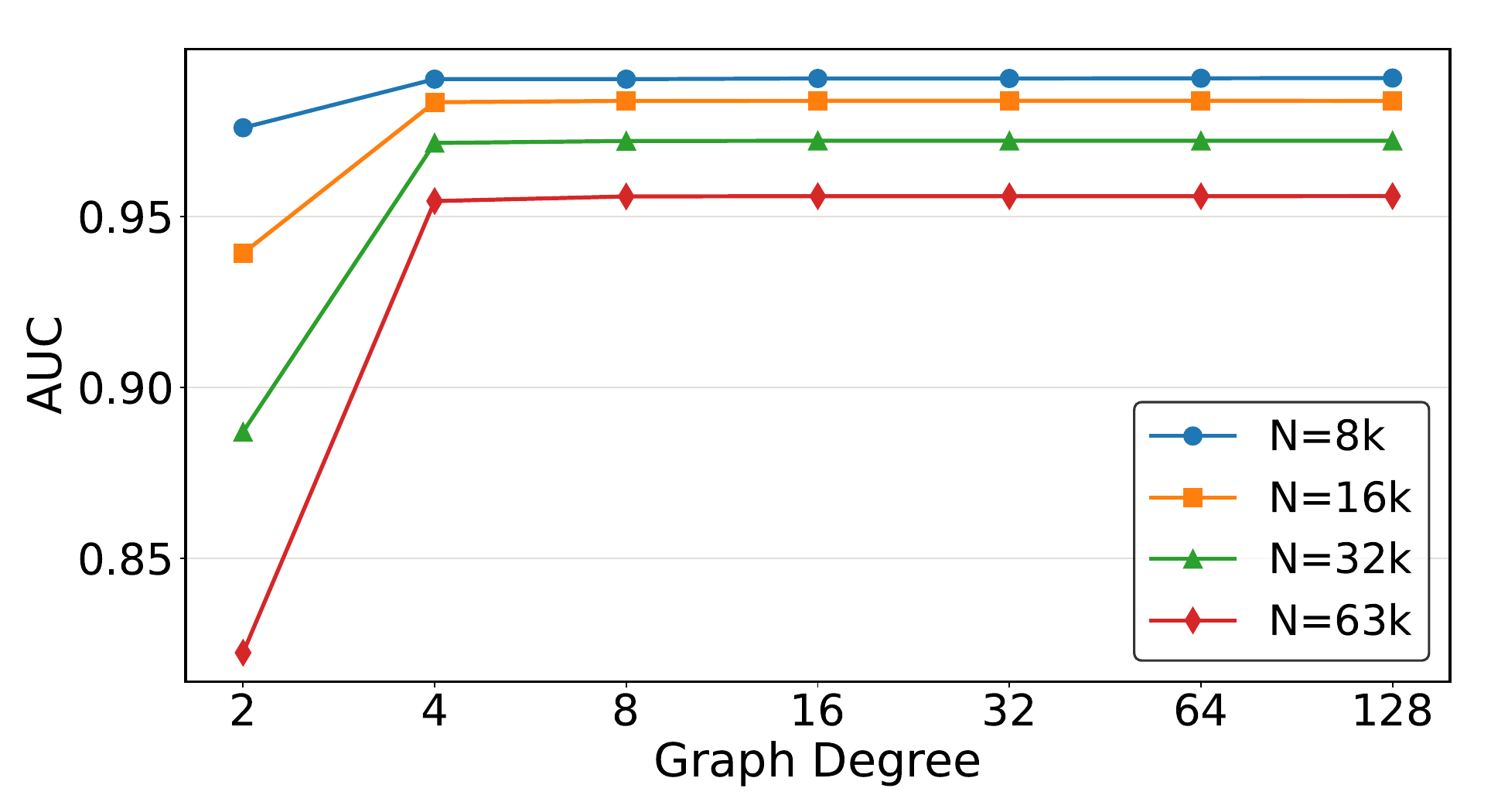}
        \caption{AUC as $N$ grows}
        \label{subfig:gd_sweep_time}
    \end{subfigure}
    \begin{subfigure}[b]{0.49\textwidth}
        \includegraphics[width=\linewidth]{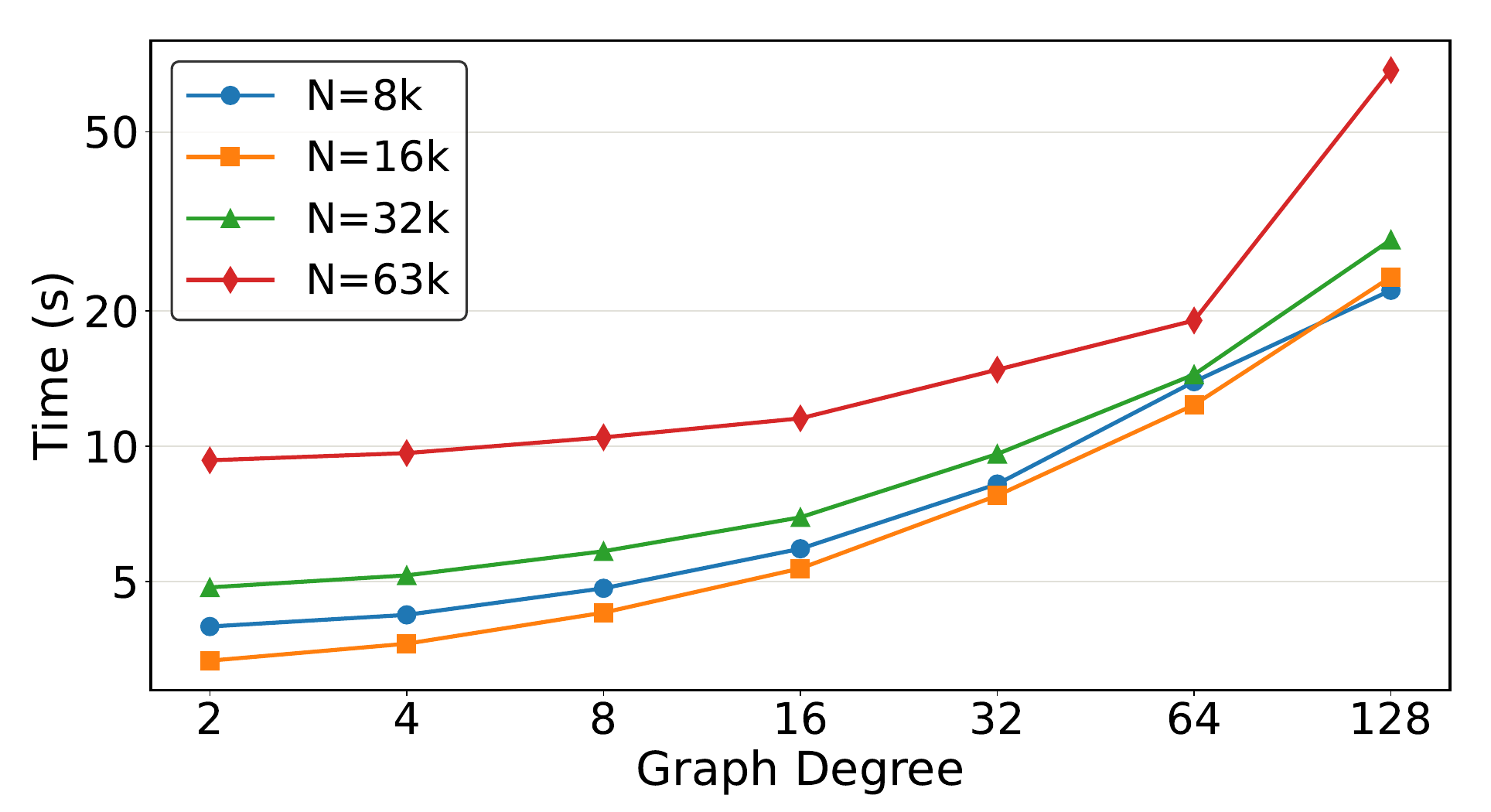}
        \caption{Correlation time}
        \label{subfig:gd_sweep_auc}
    \end{subfigure}
    \caption{Hyperparameter tuning on graph degrees}
    \label{fig:gd_sweep}
\end{figure}

Figure~\ref{fig:gd_sweep} shows detection performance (AUC) and correlation time as the graph degree $d$ varies over $\{2, 4, 8, 16, 32, 64, 128\}$ for $N \in \{8k, 16k, 32k, 63k\}$. 
AUC saturates beyond $d=4$ and does not improve further, while correlation time continues to grow as $d$ increases. We therefore choose $d=8$, retaining a small margin above the saturation point, as our default hyperparameter.

\subsection{Search width and Max iteration}
$I_{max}$ is not an independent hyperparameter: by CAGRA's implementation, it is determined by itopk\_size and $p$ as $I_{max}$ = 1 + min(1.1$x$, $x$+10) where $x$ =itopk\_size / $p$.
Table~\ref{tab:cagra_width_itopk_sweep} therefore reports the sweep over $p$ and itopk\_size directly, along with the resulting AUC, PR-AUC, and search time. 
PR-AUC degrades noticeably as itopk\_size decreases from 128 to 32, while when $p$ increases, AUC or PR-AUC improve. Search time decreases modestly as $p$ increases.
We select $p=1$ and itopk\_size = 128, yielding $I_{max}=139$.

\begin{table}[H]
\centering
\caption{CAGRA $p$ and $I_{max}$ sweep. (best per column in bold)}
\label{tab:cagra_width_itopk_sweep}
\setlength{\tabcolsep}{8pt} 
\begin{tabular}{cccccc}
\toprule
$p$ & itopk\_size & $I_{max}$ & AUC & PR-AUC & search\_time (s) \\
\midrule
1 & 128 & 139.0 & \textbf{0.9972} & \textbf{0.9271} & 0.3096 \\
2 & 128 & 71.4 & 0.9971 & \textbf{0.9271} & 0.3094 \\
4 & 128 & 36.2 & 0.9971 & 0.9268 & 0.2887 \\
8 & 128 & 18.6 & 0.9972 & 0.9267 & 0.2911 \\
\midrule
1 & 64 & 71.4 & 0.9971 & 0.9223 & 0.2883 \\
2 & 64 & 36.2 & 0.9971 & 0.9231 & 0.2868 \\
4 & 64 & 18.6 & 0.9971 & 0.9226 & 0.2815 \\
8 & 64 & 9.8 & 0.9972 & 0.9240 & 0.2813 \\
\midrule
1 & 32 & 36.2 & 0.9968 & 0.9062 & 0.2822 \\
2 & 32 & 18.6 & 0.9970 & 0.9097 & 0.2845 \\
4 & 32 & 9.8 & 0.9969 & 0.9132 & 0.2850 \\
8 & 32 & 5.4 & 0.9971 & 0.9167 & \textbf{0.2748} \\
\bottomrule
\end{tabular}
\end{table}

\subsection{Number of Nearest Neighbors}
We sweep $k$, which determines how many nearest neighbors are retrieved and thus considered as correlation candidates, over $\{2,4,8,16 \dots1024\}$.
Table~\ref{tab:top-k-sweep} shows that PR-AUC follows an inverted-U pattern: too small $k$ risks excluding the true match from the candidate set, while too large $k$ admits excessive false candidates, degrading precision. PR-AUC peaks at $k=32$ (0.9270).

\begin{table}[H]
    \centering
    \caption{PR-AUC across $k$ (best PR-AUC in bold)}
    \label{tab:top-k-sweep}
    \begin{tabular}{lcccccccccc}
        \toprule
        $k$          & 2      & 4      & 8      & 16     & 32     & 64     & 128    & 256    & 512    & 1024   \\
        \midrule
        PR-AUC       & 0.8914 & 0.9060 & 0.9177 & 0.9252 & \textbf{0.9270} & 0.9222 & 0.9153 & 0.8881 & 0.8207 & 0.6607 \\
        \bottomrule
    \end{tabular}
\end{table}

\section{Extended Time Complexity Analysis}
\label{appendix:extended_time_complexity}

\begin{figure}[H]
    \centering
    \includegraphics[width=0.5\linewidth]{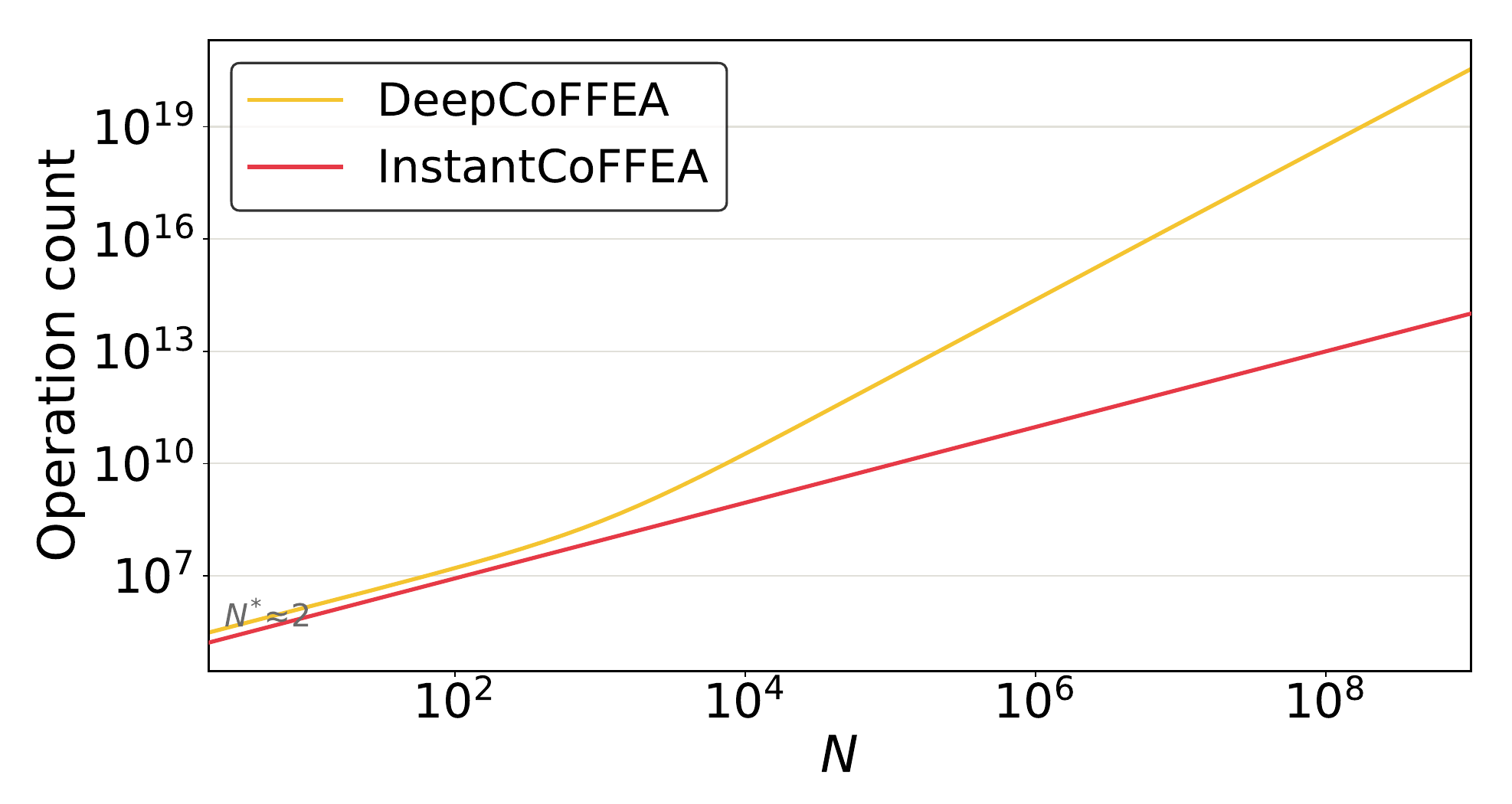}
    \caption{Number of operations where $k=1024, d=16, p=1, I_{max}=139$. Crossover point is $N=2$.}
    \label{fig:time_complexity}
\end{figure}

As shown in Table~\ref{tab:time_complexity}, the constants $W$, $k$, $d$, $p$, $I_{max}$ appear as coefficients in the time-complexity expressions, and their concrete values can affect whether the asymptotic advantage holds in practice. Figure~\ref{fig:time_complexity} shows the operation counts of DeepCoFFEA and InstantCoFFEA under the worst-case configuration of these coefficients. Even in this setting, InstantCoFFEA's operation count falls below DeepCoFFEA's once $N$ exceeds 2, and the gap widens as $N$ increases further.


\FloatBarrier

\section{Wall-Clock Time Comparison (Real vs. Jittered)}
\label{appendix:jitter_vs._real}

Table~\ref{tab:jitter_vs_real} compares InstantCoFFEA's wall-clock time on real data against duplicated and jittered data. 
We observe that wall-clock time tends to be slightly larger on real data across the tested values of $N$, which suggests that wall-clock time measured on duplicated and jittered data may somewhat underestimate the wall-clock time on real data. 
We note this as a potential limitation of using duplicated and jittered data as a proxy for benchmarking purposes.

\begin{table}[h]
\centering
\caption{InstantCoFFEA wall-clock time comparison between real and duplicated+jittered data by sample size $N$ (Ratio = Jittered/Real)}
\small
\begin{tabular}{rrrr}
\toprule
$N$ & Real (s) & Jittered (s) & Ratio \\
\midrule
7,759   & 2.720$\pm$0.771  & 2.640$\pm$0.854  & 0.97x \\
15,866  & 5.877$\pm$0.402  & 5.575$\pm$0.402  & 0.95x \\
24,210  & 7.331$\pm$0.702  & 6.707$\pm$0.309  & 0.92x \\
62,727  & 14.885$\pm$1.822 & 13.442$\pm$0.839 & 0.90x \\
\bottomrule
\end{tabular}
\label{tab:jitter_vs_real}
\end{table}

\section{Detection performance as N grows}
\label{appenix:prauc_as_N_rows}

Table~\ref{tab:prauc_as_n} compares InstantCoFFEA and DeepCoFFEA's PR-AUC across increasing pool size $N$. 
Both methods exhibit a consistent decline in PR-AUC as $N$ increases: InstantCoFFEA drops from 0.9315 at $N=7{,}166$ to 0.8397 at $N=64{,}882$, while DeepCoFFEA drops from 0.8999 to 0.7992 over the same range. 
InstantCoFFEA maintains a consistent margin over DeepCoFFEA across all tested values of $N$, indicating that InstantCoFFEA's advantage is not an artifact of a particular pool size but persists as the pool scales.

\begin{table}[h]
\centering
\caption{InstantCoFFEA and DeepCoFFEA's PR-AUC as a function of pool size $N$}
\small
\begin{tabular}{lrrrrr}
\toprule
$N$ & 7,166 & 15,518 & 31,732 & 48,420 & 64,882 \\
\midrule
InstantCoFFEA & 0.9315 & 0.9177 & 0.8837 & 0.8585 & 0.8397 \\
DeepCoFFEA    & 0.8999 & 0.8910 & 0.8490 & 0.8234 & 0.7992 \\
\bottomrule
\end{tabular}
\label{tab:prauc_as_n}
\end{table}


\section{Top-k Matching Pair Count}
\label{appendix:top_k_count}


To understand why ranked voting outperforms DeepCoFFEA despite relying on approximate (ANN-based) retrieval, we examine where true matching pairs rank among candidates and how many are captured at a given false-positive rate.

Table~\ref{tab:dcf-rank-distribution} shows the rank distribution of DeepCoFFEA's true matching pairs. 
In the lab setting, matches are concentrated at low ranks—over half fall within rank 16. 
In the realistic setting, however, this concentration disappears: nearly a third of true matches fall beyond rank 257, meaning a substantial fraction of correct pairs are pushed toward the tail of the candidate list precisely in the setting we care about most.

Despite this less favorable rank distribution, Table~\ref{tab:dcf-ours-matched} shows that ranked voting still captures more true matches than DeepCoFFEA at nearly every false-positive rate in the realistic setting (e.g., 98 vs. 75 at FPR = $10^{-8}$), while binary voting largely trails both. This indicates that ranked voting's advantage does not come from an easier ranking problem, but from making better use of the rank information itself—extracting signal that binary top-$k$ voting and DeepCoFFEA's exhaustive comparison both discard.

\begin{table}[h]
\centering
\caption{Recall@$k$ comparison between DeepCoFFEA(DCF) and \sys(ICF)}
\label{tab:recall_at_k_dcf_icf}
\begin{tabular}{llccccccccc}
\toprule
Setting & Method & \multicolumn{9}{c}{Recall@$k$} \\
\cmidrule(lr){3-11}
 & & 2 & 4 & 8 & 16 & 32 & 64 & 128 & 256 & 1024 \\
\midrule
\multirow{2}{*}{Lab setting}
 & ICF      & 0.3342 & 0.4102 & 0.4851 & 0.5623 & 0.6312 & 0.6952 & 0.7553 & 0.8254 & 0.8994 \\
 & DCF & 0.3382 & 0.4148 & 0.4906 & 0.5693 & 0.6417 & 0.7103 & 0.7762 & 0.8316 & 0.9183 \\
\midrule
\multirow{2}{*}{Realistic setting}
 & ICF      & 0.5330 & 0.6223 & 0.7019 & 0.7724 & 0.8310 & 0.8791 & 0.9165 & 0.9477 & 0.9803 \\
 & DCF & 0.5334 & 0.6228 & 0.7028 & 0.7735 & 0.8332 & 0.8818 & 0.9218 & 0.9512 & 0.9845 \\
\bottomrule
\end{tabular}
\end{table}


\begin{table}[h]
\centering
\caption{DeepCoFFEA's rank distribution of the true matching pair. Lab setting has in total 7804 matching pairs, and realistic setting has 352 pairs.}
\footnotesize
\resizebox{\textwidth}{!}{%
\begin{tabular}{lcccccc}
\toprule
Rank & 1 & 2--4 & 5--16 & 17--64 & 65--256 & 257+ \\
\midrule
Lab setting & 1323 (17.0\%) & 1400 (17.9\%) & 1781 (22.8\%) & 1690 (21.7\%) & 1118 (14.3\%) & 492 (6.3\%) \\
Realistic setting & 21 (6.0\%) & 21 (6.0\%) & 53 (15.1\%) & 70 (19.9\%) & 75 (21.3\%) & 112 (31.8\%) \\
\bottomrule
\end{tabular}%
}

\label{tab:dcf-rank-distribution}
\end{table}

\begin{table}[H]
\centering
\caption{DeepCoFFEA / InstantCoFFEA(binary voting) / InstantCoFFEA(ranked voting)'s number of true pairs exceeding the threshold at each FPR (best per column in bold) }
\scriptsize
\setlength{\tabcolsep}{4.5pt}
\begin{tabular}{lcccccccccc}
\toprule
\multicolumn{9}{l}{\textbf{Lab setting (number of matching pairs=7804)}} \\
\midrule
FPR & 7e-07 & 2e-06 & 4e-06 & 1e-05 & 3e-05 & 9e-05 & 4e-04 & 2e-03  \\
\midrule
DeepCoFFEA & 3993 & 4803 & 5615 & 6336 & 6916 & 7323 & 7565 & \textbf{7705}  \\
InstantCoFFEA (binary voting) & 2933 & 4469 & 5496 & 6261 & 6812 & 7175 & 7417 & 7680  \\
InstantCoFFEA (ranked voting) & \textbf{4452} & \textbf{5532} & \textbf{6340} & \textbf{6879} & \textbf{7169} &\textbf{7401} & \textbf{7578} & 7681   \\
\midrule
\multicolumn{9}{l}{\textbf{Realistic setting (number of matching pairs=352)}} \\
\midrule
FPR & 1e-08 & 6e-08 & 3e-07 & 8e-07 & 2e-06 & 6e-06 & 1e-05 & 4e-05  \\
\midrule
DeepCoFFEA & 75 & 112 & 146 & 181 & 225 & 258 & \textbf{285} & \textbf{309} \\
InstantCoFFEA (binary voting) & 42 & 90 & 137 & 179 & 241 & 265 & 265 & 291 \\
InstantCoFFEA (ranked voting) & \textbf{98} & \textbf{132} & \textbf{191} & \textbf{231} & \textbf{257} & \textbf{269} & 284 & 299 \\
\bottomrule
\end{tabular}

\label{tab:dcf-ours-matched}
\end{table}

\section{Voting Scheme}
\label{appendix:voting_scheme}

\begin{figure}[H]
    \centering
    \begin{subfigure}[b]{0.49\textwidth}
        \includegraphics[width=\linewidth]{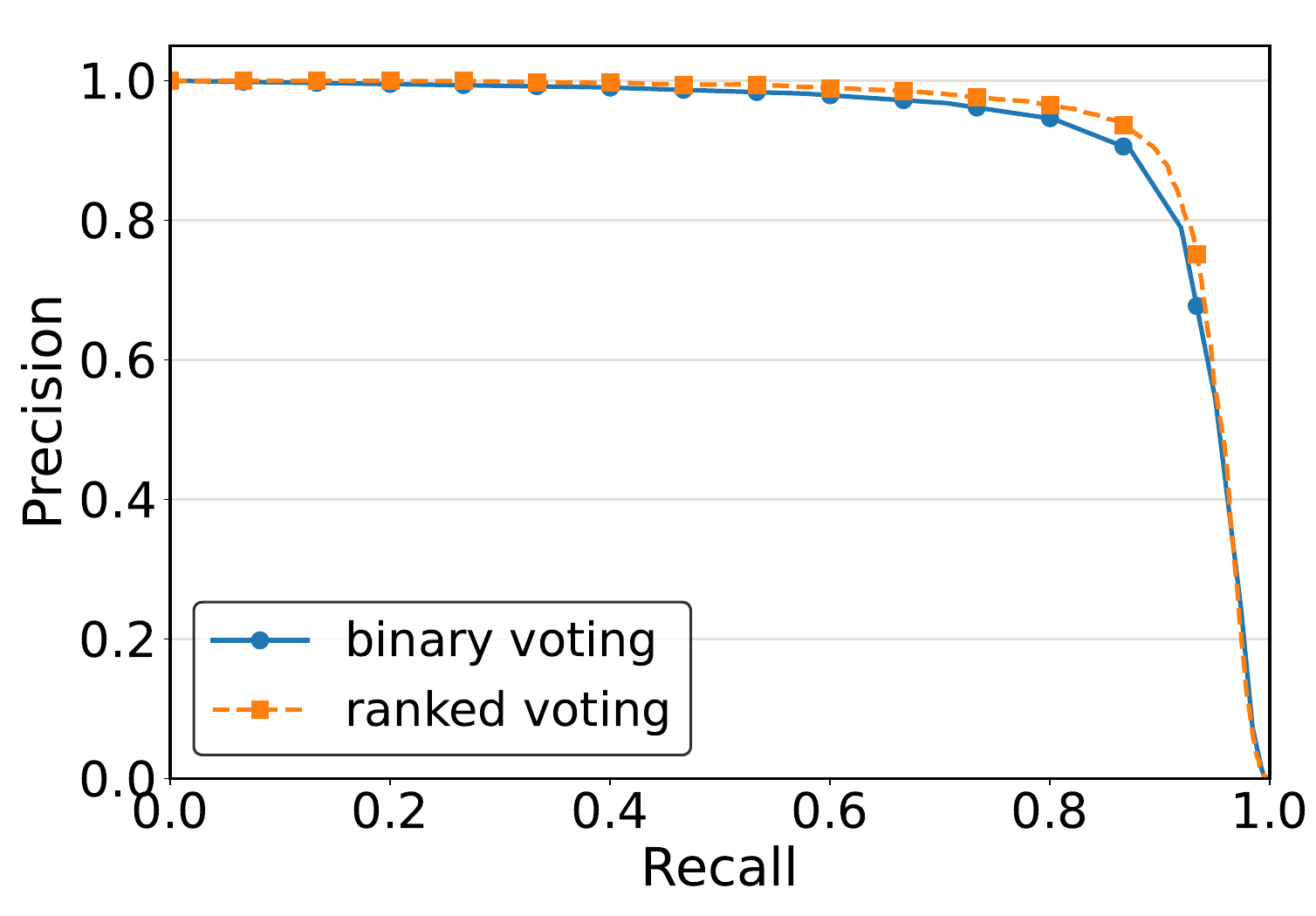}
        \caption{Lab setting}
        \label{subfig:voting_scheme_lab}
    \end{subfigure}
    \begin{subfigure}[b]{0.49\textwidth}
        \includegraphics[width=\linewidth]{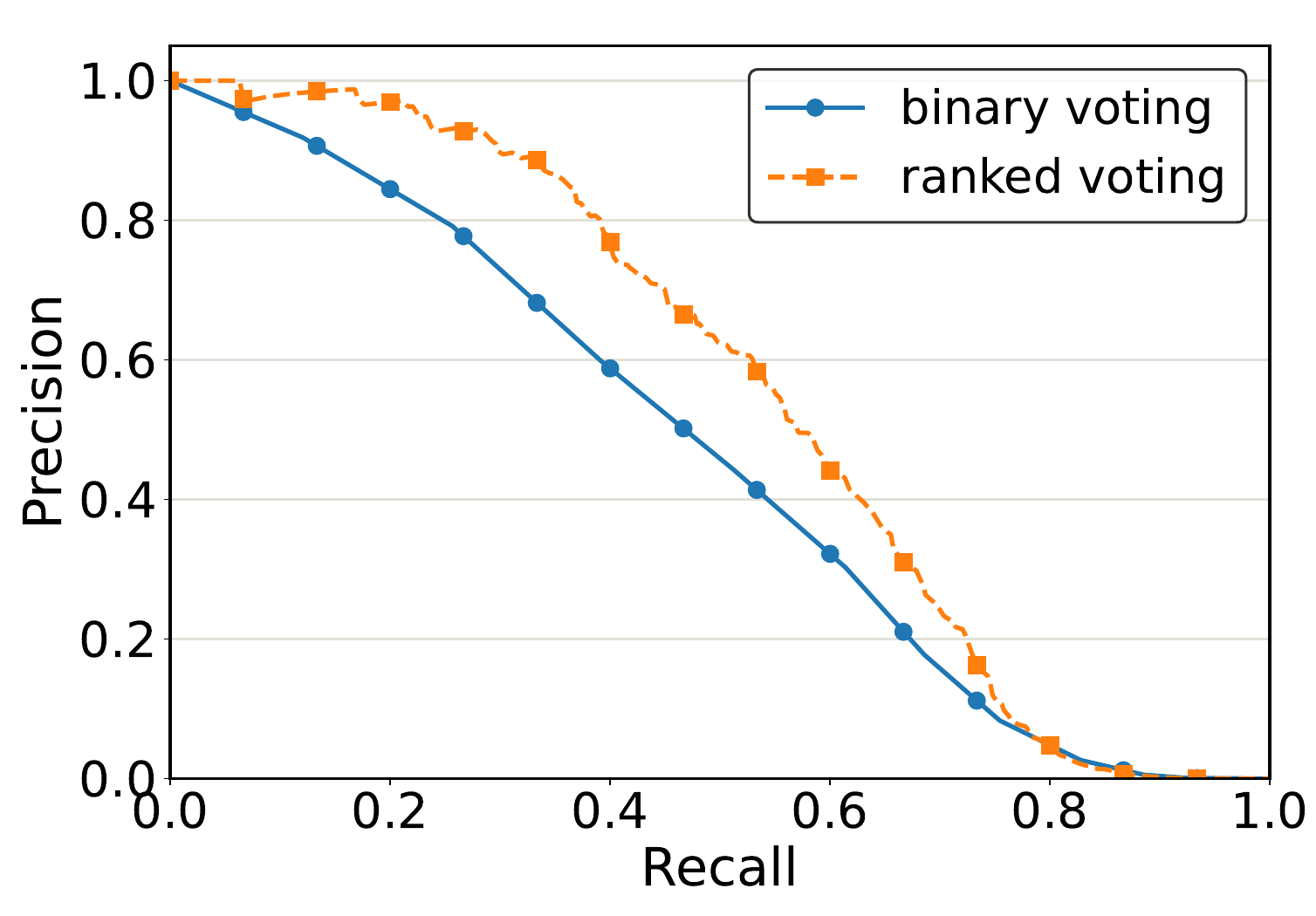}
        \caption{Realistic setting}
        \label{subfig:voting_scheme_real}
    \end{subfigure}
    \caption{PR Curve of binary voting vs. ranked voting}
    \label{fig:voting_scheme}
\end{figure}

Figure~\ref{fig:voting_scheme} shows the precision-recall curve of binary voting and ranked voting under both the lab setting and the realistic setting.
Ranked voting outperforms binary voting in both settings, but the gap is modest in the lab setting and widens substantially in the realistic setting.

\section{Network Adversary Scale Sweep Results}
\label{appendix:network_adv}

Table~\ref{tab:network-adv-all} extends the top-1 result reported in Section~\ref{sec:evaluation} to the top-5 entities across AS, IXP, and IXP Org categories. 
Due to space constraints, we report only entity IDs in Table~\ref{tab:network-adv-all}. 
The corresponding names are provided in Table~\ref{tab:ixp-id-name}.

\begin{table}[h]
\centering
\scriptsize
\setlength{\tabcolsep}{4pt}
\caption{ID and name of the top-5 IXPs and IXP organizations,
with their rank by compromise rate at 1.4M scale under each destination model.}
\label{tab:ixp-id-name}
\begin{tabular}{r l cc}
\toprule
ID & Name & Per-country & Global \\
\midrule

\multicolumn{4}{l}{\emph{Internet Exchange Point}} \\
23   & Amsterdam Internet Exchange                             & 1 & 2 \\
419  & LINX LON1                                               & 2 & 1 \\
207  & Deutscher Commercial Internet Exchange DE-CIX Frankfurt & 3 & 3 \\
576  & Neutral Internet Exchange                               & 4 & 4 \\
545  & NL-IX                                                   & 5 & 5 \\
\midrule

\multicolumn{4}{l}{\emph{IXP Organization}} \\
2634 & Amsterdam Internet Exchange B.V. & 1 & 2 \\
791  & LINX                             & 2 & 1 \\
1187 & DE-CIX Group AG                  & 3 & 3 \\
2    & Equinix, Inc.                    & 4 & 4 \\
138  & NL-ix                            & 5 & 5 \\
\bottomrule
\end{tabular}
\end{table}

\FloatBarrier   

\begin{center}
\scriptsize
\renewcommand{\arraystretch}{0.85}
\setlength{\tabcolsep}{2.5pt}
\begin{longtable}{llllrrr}
\caption{Top-5 guard/exit flow and compromise percentage, by total circuit count (Scale), and client destination selection model (Dest.: global vs. per-country) per AS, IXP, and IXP Org.} \label{tab:network-adv-all} \\
\toprule
Scale & Category & Dest.& ID & \multicolumn{1}{c}{Guard} & \multicolumn{1}{c}{Exit} & \multicolumn{1}{c}{Compr.\%} \\
\midrule
\endfirsthead
\multicolumn{7}{l}{\scriptsize\itshape (continued from previous page)} \\
\toprule
Scale & Category & Dest. & ID & \multicolumn{1}{c}{Guard} & \multicolumn{1}{c}{Exit} & \multicolumn{1}{c}{Compr.\%} \\
\midrule
\endhead
\midrule
\multicolumn{7}{r}{\scriptsize\itshape (continued on next page)} \\
\endfoot
\bottomrule
\endlastfoot
1.4M & AS & global & 174 & 255,968$\pm$6,082 & 186,358$\pm$3,756 & 2.903$\pm$0.041 \\
 &  &  & 6453 & 185,377$\pm$4,458 & 174,457$\pm$3,979 & 1.990$\pm$0.011 \\
 &  &  & 9790 & 133,446$\pm$3,217 & 56,806$\pm$822 & 0.452$\pm$0.006 \\
 &  &  & 1299 & 139,589$\pm$3,317 & 52,103$\pm$1,225 & 0.443$\pm$0.009 \\
 &  &  & 6939 & 127,159$\pm$2,835 & 41,660$\pm$1,338 & 0.326$\pm$0.006 \\
\cmidrule(lr){4-7}
 & AS & per-country & 6453 & 151,232$\pm$3,588 & 224,224$\pm$6,896 & 2.082$\pm$0.019 \\
 &  &  & 174 & 227,097$\pm$5,286 & 115,126$\pm$2,218 & 1.945$\pm$0.029 \\
 &  &  & 16276 & 206,891$\pm$4,938 & 20,036$\pm$467 & 0.957$\pm$0.009 \\
 &  &  & 9790 & 111,788$\pm$2,641 & 93,508$\pm$1,691 & 0.802$\pm$0.009 \\
 &  &  & 1299 & 125,844$\pm$2,950 & 61,275$\pm$1,361 & 0.555$\pm$0.012 \\
\cmidrule(lr){4-7}
 & IXP & global & 419 & 389,682$\pm$9,160 & 657,795$\pm$18,045 & 16.004$\pm$0.097 \\
 &  &  & 23 & 315,495$\pm$7,306 & 774,979$\pm$22,968 & 15.088$\pm$0.109 \\
 &  &  & 207 & 324,750$\pm$7,538 & 326,409$\pm$4,905 & 6.586$\pm$0.107 \\
 &  &  & 576 & 92,490$\pm$2,128 & 507,707$\pm$16,862 & 2.898$\pm$0.033 \\
 &  &  & 545 & 89,156$\pm$2,062 & 504,828$\pm$16,844 & 2.776$\pm$0.032 \\
\cmidrule(lr){4-7}
 & IXP & per-country & 23 & 282,661$\pm$6,448 & 657,497$\pm$19,472 & 12.913$\pm$0.095 \\
 &  &  & 419 & 342,885$\pm$7,973 & 410,234$\pm$10,502 & 10.600$\pm$0.049 \\
 &  &  & 207 & 280,603$\pm$6,447 & 236,118$\pm$3,897 & 4.921$\pm$0.065 \\
 &  &  & 576 & 81,974$\pm$1,850 & 315,781$\pm$9,241 & 1.912$\pm$0.019 \\
 &  &  & 545 & 79,356$\pm$1,792 & 313,146$\pm$9,214 & 1.829$\pm$0.019 \\
\cmidrule(lr){4-7}
 & IXP Org & global & 791 & 390,032$\pm$9,164 & 658,017$\pm$18,049 & 16.024$\pm$0.097 \\
 &  &  & 2634 & 315,554$\pm$7,307 & 775,000$\pm$22,967 & 15.091$\pm$0.109 \\
 &  &  & 1187 & 342,421$\pm$7,949 & 331,964$\pm$4,962 & 7.062$\pm$0.116 \\
 &  &  & 2 & 240,110$\pm$5,599 & 303,538$\pm$4,108 & 4.538$\pm$0.079 \\
 &  &  & 138 & 92,490$\pm$2,128 & 507,707$\pm$16,862 & 2.898$\pm$0.033 \\
\cmidrule(lr){4-7}
 & IXP Org & per-country & 2634 & 282,726$\pm$6,449 & 657,503$\pm$19,472 & 12.916$\pm$0.095 \\
 &  &  & 791 & 343,276$\pm$7,981 & 410,234$\pm$10,502 & 10.612$\pm$0.049 \\
 &  &  & 1187 & 298,297$\pm$6,849 & 241,494$\pm$4,060 & 5.339$\pm$0.070 \\
 &  &  & 2 & 226,506$\pm$5,246 & 260,548$\pm$3,872 & 4.449$\pm$0.055 \\
 &  &  & 138 & 81,974$\pm$1,850 & 315,781$\pm$9,241 & 1.912$\pm$0.019 \\
\midrule
3M & AS & global & 174 & 547,391$\pm$13,052 & 399,120$\pm$7,719 & 2.901$\pm$0.049 \\
 &  &  & 6453 & 397,095$\pm$9,158 & 373,553$\pm$8,663 & 1.983$\pm$0.009 \\
 &  &  & 9790 & 285,712$\pm$6,940 & 121,760$\pm$1,784 & 0.450$\pm$0.006 \\
 &  &  & 1299 & 298,796$\pm$7,192 & 111,545$\pm$2,577 & 0.443$\pm$0.009 \\
 &  &  & 6939 & 273,340$\pm$6,172 & 89,391$\pm$2,875 & 0.327$\pm$0.006 \\
\cmidrule(lr){4-7}
 & AS & per-country & 6453 & 323,768$\pm$7,477 & 480,748$\pm$14,729 & 2.082$\pm$0.019 \\
 &  &  & 174 & 485,340$\pm$11,380 & 246,591$\pm$4,367 & 1.941$\pm$0.030 \\
 &  &  & 16276 & 443,388$\pm$10,530 & 42,868$\pm$947 & 0.957$\pm$0.005 \\
 &  &  & 9790 & 239,249$\pm$5,819 & 200,250$\pm$3,618 & 0.802$\pm$0.005 \\
 &  &  & 1299 & 269,496$\pm$6,407 & 131,098$\pm$2,877 & 0.551$\pm$0.012 \\
\cmidrule(lr){4-7}
 & IXP & global & 419 & 835,244$\pm$19,688 & 1,409,575$\pm$38,885 & 16.011$\pm$0.101 \\
 &  &  & 23 & 676,742$\pm$15,401 & 1,660,936$\pm$49,085 & 15.101$\pm$0.104 \\
 &  &  & 207 & 696,622$\pm$16,105 & 699,720$\pm$10,657 & 6.594$\pm$0.115 \\
 &  &  & 576 & 198,488$\pm$4,516 & 1,088,324$\pm$35,950 & 2.898$\pm$0.034 \\
 &  &  & 545 & 191,211$\pm$4,379 & 1,082,132$\pm$35,862 & 2.774$\pm$0.034 \\
\cmidrule(lr){4-7}
 & IXP & per-country & 23 & 606,431$\pm$13,743 & 1,409,261$\pm$42,312 & 12.928$\pm$0.089 \\
 &  &  & 419 & 735,257$\pm$17,225 & 879,216$\pm$23,016 & 10.612$\pm$0.045 \\
 &  &  & 207 & 602,106$\pm$13,871 & 506,358$\pm$8,917 & 4.928$\pm$0.065 \\
 &  &  & 576 & 175,920$\pm$3,997 & 676,668$\pm$19,924 & 1.911$\pm$0.019 \\
 &  &  & 545 & 170,212$\pm$3,894 & 671,021$\pm$19,851 & 1.826$\pm$0.018 \\
\cmidrule(lr){4-7}
 & IXP Org & global & 791 & 836,008$\pm$19,699 & 1,410,044$\pm$38,887 & 16.031$\pm$0.100 \\
 &  &  & 2634 & 676,850$\pm$15,404 & 1,660,981$\pm$49,088 & 15.104$\pm$0.104 \\
 &  &  & 1187 & 734,302$\pm$16,973 & 711,627$\pm$10,768 & 7.068$\pm$0.124 \\
 &  &  & 2 & 514,916$\pm$12,072 & 650,634$\pm$8,843 & 4.544$\pm$0.082 \\
 &  &  & 138 & 198,488$\pm$4,516 & 1,088,324$\pm$35,950 & 2.898$\pm$0.034 \\
\cmidrule(lr){4-7}
 & IXP Org & per-country & 2634 & 606,552$\pm$13,747 & 1,409,273$\pm$42,312 & 12.930$\pm$0.089 \\
 &  &  & 791 & 736,104$\pm$17,235 & 879,216$\pm$23,016 & 10.624$\pm$0.045 \\
 &  &  & 1187 & 639,822$\pm$14,732 & 517,925$\pm$9,240 & 5.346$\pm$0.071 \\
 &  &  & 2 & 485,858$\pm$11,398 & 558,757$\pm$8,806 & 4.460$\pm$0.051 \\
 &  &  & 138 & 175,920$\pm$3,997 & 676,668$\pm$19,924 & 1.911$\pm$0.019 \\
\midrule
6M & AS & global & 174 & 1,094,497$\pm$25,882 & 798,242$\pm$15,837 & 2.899$\pm$0.046 \\
 &  &  & 6453 & 794,685$\pm$18,684 & 747,363$\pm$17,306 & 1.987$\pm$0.008 \\
 &  &  & 9790 & 571,694$\pm$13,983 & 243,649$\pm$3,187 & 0.451$\pm$0.005 \\
 &  &  & 1299 & 598,659$\pm$14,356 & 223,297$\pm$5,347 & 0.443$\pm$0.011 \\
 &  &  & 6939 & 545,381$\pm$12,171 & 178,941$\pm$5,762 & 0.327$\pm$0.005 \\
\cmidrule(lr){4-7}
 & AS & per-country & 6453 & 647,964$\pm$15,178 & 961,519$\pm$28,880 & 2.084$\pm$0.017 \\
 &  &  & 174 & 971,085$\pm$22,824 & 493,138$\pm$9,309 & 1.943$\pm$0.027 \\
 &  &  & 16276 & 886,716$\pm$21,196 & 85,930$\pm$1,888 & 0.959$\pm$0.004 \\
 &  &  & 9790 & 478,570$\pm$11,670 & 400,694$\pm$7,030 & 0.802$\pm$0.004 \\
 &  &  & 1299 & 540,142$\pm$12,883 & 262,032$\pm$5,984 & 0.554$\pm$0.013 \\
\cmidrule(lr){4-7}
 & IXP & global & 419 & 1,670,301$\pm$39,180 & 2,819,164$\pm$77,295 & 16.003$\pm$0.095 \\
 &  &  & 23 & 1,351,497$\pm$30,871 & 3,321,301$\pm$98,252 & 15.072$\pm$0.105 \\
 &  &  & 207 & 1,392,640$\pm$32,326 & 1,399,193$\pm$20,744 & 6.593$\pm$0.110 \\
 &  &  & 576 & 396,949$\pm$9,077 & 2,175,976$\pm$71,642 & 2.901$\pm$0.034 \\
 &  &  & 545 & 382,412$\pm$8,759 & 2,163,637$\pm$71,462 & 2.777$\pm$0.033 \\
\cmidrule(lr){4-7}
 & IXP & per-country & 23 & 1,210,416$\pm$27,511 & 2,818,273$\pm$84,131 & 12.903$\pm$0.086 \\
 &  &  & 419 & 1,469,557$\pm$34,368 & 1,757,714$\pm$46,453 & 10.600$\pm$0.046 \\
 &  &  & 207 & 1,202,723$\pm$27,865 & 1,012,826$\pm$17,272 & 4.923$\pm$0.065 \\
 &  &  & 576 & 351,519$\pm$7,975 & 1,352,989$\pm$40,503 & 1.912$\pm$0.017 \\
 &  &  & 545 & 340,185$\pm$7,731 & 1,341,657$\pm$40,325 & 1.827$\pm$0.016 \\
\cmidrule(lr){4-7}
 & IXP Org & global & 791 & 1,671,807$\pm$39,204 & 2,820,104$\pm$77,309 & 16.023$\pm$0.095 \\
 &  &  & 2634 & 1,351,737$\pm$30,877 & 3,321,385$\pm$98,258 & 15.076$\pm$0.105 \\
 &  &  & 1187 & 1,468,236$\pm$34,035 & 1,423,063$\pm$20,928 & 7.068$\pm$0.119 \\
 &  &  & 2 & 1,030,043$\pm$24,102 & 1,301,378$\pm$17,190 & 4.545$\pm$0.081 \\
 &  &  & 138 & 396,949$\pm$9,077 & 2,175,976$\pm$71,642 & 2.901$\pm$0.034 \\
\cmidrule(lr){4-7}
 & IXP Org & per-country & 2634 & 1,210,688$\pm$27,518 & 2,818,299$\pm$84,131 & 12.906$\pm$0.086 \\
 &  &  & 791 & 1,471,221$\pm$34,396 & 1,757,714$\pm$46,453 & 10.612$\pm$0.046 \\
 &  &  & 1187 & 1,278,317$\pm$29,569 & 1,036,021$\pm$17,871 & 5.342$\pm$0.071 \\
 &  &  & 2 & 970,787$\pm$22,673 & 1,117,791$\pm$16,916 & 4.453$\pm$0.054 \\
 &  &  & 138 & 351,519$\pm$7,975 & 1,352,989$\pm$40,503 & 1.912$\pm$0.017 \\
\midrule
9M & AS & global & 174 & 1,639,872$\pm$38,554 & 1,195,155$\pm$23,785 & 2.901$\pm$0.047 \\
 &  &  & 6453 & 1,188,996$\pm$27,975 & 1,119,146$\pm$25,991 & 1.986$\pm$0.009 \\
 &  &  & 9790 & 855,375$\pm$20,950 & 364,754$\pm$4,882 & 0.450$\pm$0.005 \\
 &  &  & 1299 & 897,384$\pm$21,568 & 334,477$\pm$7,612 & 0.444$\pm$0.010 \\
 &  &  & 6939 & 817,920$\pm$18,172 & 267,653$\pm$8,550 & 0.326$\pm$0.005 \\
\cmidrule(lr){4-7}
 & AS & per-country & 6453 & 969,654$\pm$22,605 & 1,440,195$\pm$44,693 & 2.082$\pm$0.020 \\
 &  &  & 174 & 1,454,977$\pm$34,276 & 738,880$\pm$13,808 & 1.944$\pm$0.030 \\
 &  &  & 16276 & 1,326,984$\pm$31,506 & 128,700$\pm$2,802 & 0.959$\pm$0.004 \\
 &  &  & 9790 & 716,023$\pm$17,320 & 599,993$\pm$10,246 & 0.802$\pm$0.004 \\
 &  &  & 1299 & 809,750$\pm$19,203 & 392,720$\pm$8,717 & 0.554$\pm$0.011 \\
\cmidrule(lr){4-7}
 & IXP & global & 419 & 2,502,230$\pm$58,620 & 4,222,477$\pm$115,827 & 16.008$\pm$0.093 \\
 &  &  & 23 & 2,025,698$\pm$46,021 & 4,974,668$\pm$146,769 & 15.088$\pm$0.103 \\
 &  &  & 207 & 2,085,502$\pm$48,047 & 2,095,098$\pm$31,076 & 6.589$\pm$0.109 \\
 &  &  & 576 & 595,476$\pm$13,671 & 3,259,562$\pm$106,890 & 2.905$\pm$0.031 \\
 &  &  & 545 & 573,673$\pm$13,162 & 3,241,068$\pm$106,605 & 2.780$\pm$0.030 \\
\cmidrule(lr){4-7}
 & IXP & per-country & 23 & 1,814,440$\pm$40,992 & 4,222,067$\pm$125,959 & 12.915$\pm$0.084 \\
 &  &  & 419 & 2,201,925$\pm$51,358 & 2,633,244$\pm$69,077 & 10.608$\pm$0.044 \\
 &  &  & 207 & 1,801,906$\pm$41,489 & 1,516,647$\pm$25,860 & 4.926$\pm$0.068 \\
 &  &  & 576 & 527,782$\pm$11,948 & 2,027,274$\pm$59,969 & 1.915$\pm$0.017 \\
 &  &  & 545 & 510,682$\pm$11,539 & 2,010,330$\pm$59,706 & 1.830$\pm$0.016 \\
\cmidrule(lr){4-7}
 & IXP Org & global & 791 & 2,504,457$\pm$58,690 & 4,223,889$\pm$115,853 & 16.028$\pm$0.093 \\
 &  &  & 2634 & 2,026,052$\pm$46,031 & 4,974,796$\pm$146,778 & 15.091$\pm$0.103 \\
 &  &  & 1187 & 2,198,776$\pm$50,644 & 2,130,857$\pm$31,491 & 7.065$\pm$0.118 \\
 &  &  & 2 & 1,540,863$\pm$36,067 & 1,948,597$\pm$26,068 & 4.538$\pm$0.079 \\
 &  &  & 138 & 595,476$\pm$13,671 & 3,259,562$\pm$106,890 & 2.905$\pm$0.031 \\
\cmidrule(lr){4-7}
 & IXP Org & per-country & 2634 & 1,814,844$\pm$41,002 & 4,222,105$\pm$125,960 & 12.918$\pm$0.084 \\
 &  &  & 791 & 2,204,386$\pm$51,432 & 2,633,244$\pm$69,077 & 10.619$\pm$0.044 \\
 &  &  & 1187 & 1,915,116$\pm$44,033 & 1,551,332$\pm$26,744 & 5.345$\pm$0.073 \\
 &  &  & 2 & 1,452,930$\pm$33,796 & 1,673,672$\pm$25,399 & 4.449$\pm$0.056 \\
 &  &  & 138 & 527,782$\pm$11,948 & 2,027,274$\pm$59,969 & 1.915$\pm$0.017 \\
\end{longtable}
\end{center}

\section{Memory usage}
\label{appendix:memory_spec}

Table~\ref{tab:memory_usage} reports the peak GPU and host memory usage of \sys and DeepCoFFEA on an NVIDIA RTX A6000 with 48GB memory. 

\begin{table*}[h]
\centering
\small
\renewcommand{\arraystretch}{0.9}
\setlength{\tabcolsep}{5pt}
\caption{Memory usage (GB) of \sys and DeepCoFFEA. OOM denotes an out-of-memory failure.}
\label{tab:memory_usage}
\begin{tabular}{l rrrr @{\hspace{18pt}} l rrrr}
\toprule
 & \multicolumn{2}{c}{\sys} & \multicolumn{2}{c}{DeepCoFFEA} &
 & \multicolumn{2}{c}{\sys} & \multicolumn{2}{c}{DeepCoFFEA} \\
\cmidrule(lr){2-3}\cmidrule(lr){4-5} \cmidrule(lr){7-8}\cmidrule(lr){9-10}
Scale & GPU & Host & GPU & Host & Scale & GPU & Host & GPU & Host \\
\midrule
\multicolumn{5}{@{}l}{\textbf{2{,}000\,MB/s relay}} & \multicolumn{5}{l}{\textbf{4{,}000\,MB/s relay}} \\
1.4M & 0.39 & 2.64  & 3.96  & 5.65  & 1.4M & 0.40 & 3.67  & 6.16  & 6.46  \\
3M & 0.48 & 4.36  & 7.62  & 7.02  & 3M & 0.58 & 6.66  & 12.39 & 8.90  \\
6M   & 0.57 & 5.93  & 10.79 & 8.31  & 6M   & 0.67 & 10.07 & 19.60 & 11.70 \\
9M   & 0.63 & 8.07  & 15.40 & 10.12 & 9M   & 0.74 & 14.47 & 28.55 & 15.13 \\
\midrule
\multicolumn{5}{@{}l}{\textbf{AS174}} & \multicolumn{5}{l}{\textbf{AMS-IX}} \\
1.4M & 0.58 & 11.23 & 13.63 & 12.61 & 1.4M & 0.87 & 22.77  & 44.05 & 22.97 \\
3M   & 0.83 & 22.92 & 27.38 & 23.12 & 3M   & 1.36 & 48.32  & \multicolumn{2}{c}{OOM} \\
6M   & 1.32 & 45.22 & \multicolumn{2}{c}{OOM} & 6M   & 2.30 & 95.94  & \multicolumn{2}{c}{OOM} \\
9M   & 1.81 & 67.70 & \multicolumn{2}{c}{OOM} & 9M   & 3.25 & 143.42 & \multicolumn{2}{c}{OOM} \\
\bottomrule
\end{tabular}
\end{table*}

\end{document}